\documentclass[review,3p]{elsarticle}

\usepackage[numbers]{natbib}
\usepackage{mathtools}
\usepackage{algorithmic}
\usepackage{xr}
\usepackage{datatool}
\usepackage{steinmetz} %Phase symbol for Algo. 1
\usepackage{graphicx}
\usepackage{url}
\usepackage{eso-pic}
\usepackage{comment}
\usepackage{pdfpages}
\usepackage{float}
\usepackage{svg}
\usepackage{array}
\usepackage{textcomp}
\usepackage{upgreek}
\usepackage{multirow}
\usepackage{color}
\definecolor{gray97}{gray}{.97}
\definecolor{gray75}{gray}{.75}
\definecolor{gray45}{gray}{.45}
\usepackage{etoolbox}
\usepackage{natbib}
\usepackage{amssymb}

\DeclarePairedDelimiter\floor{\lfloor}{\rfloor}

\makeatletter
\newcommand*{\addFileDependency}[1]{
	\typeout{(#1)}
	\@addtofilelist{#1}
	\IfFileExists{#1}{}{\typeout{No file #1.}}
}
\makeatother

\newcommand*{\myexternaldocument}[1]{
	\externaldocument{#1}
	\addFileDependency{#1.tex}
	\addFileDependency{#1.aux}
}

\myexternaldocument{supplemental}

\DTLnewdb{bibnotes}

\makeatletter
\patchcmd{\@lbibitem}%
{\item[\hfil\NAT@anchor{#2}{\NAT@num}]}%
{%
	\item[\hfil\NAT@anchor{#2}{\NAT@num}]%
	\DTLforeach[\DTLiseq{\mylabel}{#2}]{bibnotes}{\mylabel=mylabel,\mynote=mynote}{\textit{\mynote}}
}{}{\message{^^JPatching failed^^J}}%

\makeatletter
\newcommand\blfootnote[1]{%
	\begingroup
	\renewcommand\thefootnote{}\footnote{#1}%
	\addtocounter{footnote}{-1}%
	\endgroup
}
\makeatother
\journal{Signal Processing}

\begin{document}

\title{Wave-shape Function Model Order Estimation by Trigonometric Regression}

\author[1]{Joaquin Ruiz\corref{cor1}}

\ead{jruiz@ingenieria.uner.edu.ar}

\author[1]{Marcelo A. Colominas}
\ead{macolominas@conicet.gov.ar}
\ead{ORCID(s): 0000-0002-0201-2190 (J. Ruiz); 0000-0003-4418-7527 (M.A. Colominas)}

\cortext[cor1]{Corresponding author}
\address[1]{Institute for Research and Development in Bioengineering and Bioinformatics (IBB), CONICET-UNER, Ruta Prov. 11 Km. 10, E3100, Oro Verde, Entre R\'ios, Argentina}

\begin{abstract}
	The adaptive non-harmonic (ANH) model is a powerful tool to compactly represent oscillating signals with time-varying amplitude and phase, and non-sinusoidal oscillating morphology. Given good estimators of instantaneous amplitude and phase we can construct an adaptive model, where the morphology of the oscillation is described by the wave-shape function (WSF), a 2$\pi$-periodic more general periodic function. In this paper, we address the problem of estimating the number of harmonic components of the WSF, a problem that remains underresearched, by adapting trigonometric regression model selection criteria into this context. We study the application of these criteria, originally developed in the context of stationary signals, to the case of signals with time-varying amplitudes and phases. We then incorporate the order estimation to the ANH model reconstruction procedure and analyze its performance for noisy AM-FM signals. Experimental results on synthethic signals indicate that these criteria enable the adaptive estimation of the waveform of non-stationary signals with non-sinusoidal oscillatory patterns, even in the presence of considerable amount of noise. We also apply our reconstruction procedure to the task of denoising simulated pulse wave signals and determine that the proposed technique performs competitively to other denoising schemes. We conclude this work by showing that our adaptive order estimation algorithm takes into account the interpatient waveform variability of the electrocardiogram (ECG) and respiratory signals by analyzing recordings from the Fantasia Database. 
\end{abstract}

\maketitle
\section{Introduction}
\label{sec:Introduction}
\blfootnote{Article accepted for publication on March 18, 2022. DOI: \url{https://doi.org/10.1016/j.sigpro.2022.108543}}\blfootnote{\copyright\ 2022. This manuscript version is made available under the CC-BY-NC-ND 4.0 license: \\https://creativecommons.org/licenses/by-nc-nd/4.0/} Amplitude-modulated and frequency-modulated (AM-FM) signals are commonly encountered in natural and artificial systems~\cite{flandrin2001}. Time-varying quantities are studied in different scientific fields such as physics, electronics, medicine, or finance. This type of signals can be thought of as the output of systems whose properties change with time. These are called \emph{dynamical systems} and the signals they generate are known as \emph{non-stationary}. This non-stationarity condition poses a challenge for the analysis of the temporal and frequential properties of these signals. Classical tools like the Fourier transform (FT) and the periodogram cannot capture the time-varying patterns in the frequency and amplitude of AM-FM signals. In order to overcome this limitation, time-frequency (TF) analysis tools have been developed and used for years to study non-stationary oscillatory signals, e.g., the short-time Fourier transform (STFT), wavelet transform (WT), quadratic time-frequency distributions (QTFD), to name but a few. One of the main goals of this type of analysis is identifying the way that amplitude and frequency vary through time. These quantities are usually known as \emph{instantaneous amplitude} and \emph{instantaneous frequency}, respectively~\cite{picinbono1997instantaneous}.

Generally, an AM-FM signal can be viewed as a real-valued function $x(t)$, where the variable $t$ usually represents time. A superimposition of them constitutes a classical example to model a significant number of real-world signals, resulting in the \emph{adaptive harmonic} (AH) model: 
\begin{equation}
x(t) = \sum_{k=1}^{K} A_k(t)\cos(2\pi \phi_k(t)),
\label{eq:harmonic}
\end{equation}

\noindent with $A_k(t)$,\ $\phi_k'(t) > 0\ \forall t$. In this formula, each term of the sum is called a \emph{component} of the \emph{multicomponent} AM-FM signal $x(t)$. $A_k(t)$ and $\phi_k(t)$ are, respectively, the instantaneous amplitude and instantaneous phase of the $k$-th component. As we will soon discuss, this representation can be generalized by replacing the cosine function in each term with more general \emph{wave-shape functions} (WSFs) $s_k(\cdot)$~\cite{wu2013instantaneous}, one for each component. These WSFs allow the model to represent more complex non-harmonic oscillatory patterns that are commonplace in various fields of science, including medicine \cite{wu2014using,wu2016modeling,su2017extract}. Given estimates of $A_k(t)$, $\phi_k(t)$ and $s_k(t)$, we can model the AM-FM signal by applying the formula 

\begin{equation}
x(t) = \sum_{k=1}^{K}A_k(t)s_k(2\pi\phi_k(t)).
\label{eq:WSFK_intro}
\end{equation}

The WSF model has been successfully applied to a number of biomedical signals, including ECG~\cite{wu2014using}, respiratory~\cite{wu2014assess}, pulse wave signal~\cite{wu2016modeling} and photoplethysmography~\cite{cicone2017nonlinear}. The approach is to decompose each $s_k(\cdot)$ into a Fourier basis and the task of estimating the WSFs reduces to finding the Fourier coefficients that give the best representation of the wave-shape. These coefficients can be estimated by solving a linear regression problem on an \emph{ad hoc} pseudo-Fourier dictionary. Although this procedure for coefficients estimation is straightforward, there is a key aspect which remains unclear: how many Fourier coefficients are necessary to represent each $s_k$. In some modelling tasks, this quantity is determined based on some \emph{a priori} knowledge of the signal being analyzed. Although this \emph{ad hoc} approach can result in wave-shapes that are visually appealing, it cannot account for wave-shape variability. Therefore, it is necessary to determine the number of Fourier coefficients, i.e. the number of harmonics, automatically. A number of techniques have been presented in the literature for automatic component counting of AM-FM signals. In \cite{sucic2011estimating}, the authors propose a method for estimating the number of nonstationary components by using a Renyí entropy-based criterion. Given that the Renyí entropy criterion can only reliably estimate the number of components when there is no overlap and the components have the same time support, the authors apply a masking window over the signal and estimate the number of components within said window. Building onto this, authors of \cite{bruni2020signal} propose an alternative strategy based on binary thresholding of the spectrogram followed by Run Length Encoding (RLE) of the binarized representation. RLE is applied to each column of the binary spectrogram, the number of modes is counted and a complexity measure is derived from the difference between the number of modes from one column to the next. This approach has an improved performance when dealing with components that overlap in the T-F plane when compared to the Renyí entropy method. Both previous methods operate on a window-to-window basis over the spectrogram and require a preceding thresholding step. In the first case, the thresholding is required to eliminate noise whereas the threshold of the second method is used obtain the binarized representation. By contrast, our method does not require a previous thresholding or denoising of the signal to accurately estimate the number of components. Moreover, we show in Sec. V that our denoising approach based on the ANH model with automatic model order estimation performs competitively (and regularly outperforms) a denoising scheme based on STFT-thresholding.

Another limitation of the existing methods is that they consider each ridge as a separate sinusoidal component of the signal, whereas our model order estimation method was developed specifically for the ANH model, which models the signal as a superposition of non-sinusoidal oscillations. We will also show that our method is suitable to the case of signal components with different wave-shape functions. To the best of our knowledge, our work is the first to tackle the task of automatic estimation of model order in the context of the adaptive non-harmonic model for both monocomponent and multicomponent signals with fixed WSFs.

In this work, we propose to estimate the number of harmonic components of the Fourier expansion of the WSF by applying model selection criteria that were originally developed in the framework of trigonometric regression \cite{kavalieris1994determining}. This type of parametric regression models data by a linear combination of trigonometric functions plus a random term that usually represents noise. In this context, different criteria has been proposed to determine the number of trigonometric terms. In general, as we increase the number of model parameters we will get a better fit to the data, but when we are dealing with noisy data increasing the model size could lead to overfitting. On the other hand, if the chosen model has very few parameters we may get poor results due to bad approximation. Therefore, we need to choose the optimal number of harmonics based on some criterion. Commonly, this criterion defines a trade-off between approximation error and model size, and the best fit is obtained by selecting the parameters that minimize the criterion.

We will study the applicability of the trigonometric regression model selection criteria to the problem of estimating the optimal number of harmonics of the AM-FM WSF model. In Sec. \ref{sec:ANH}, we will give details of the adaptive harmonic model, show its limitations and introduce the non-harmonic variant proposed by HT Wu in~\cite{wu2013instantaneous}. In Sec. \ref{sec:Criteria} we will discuss the trigonometric regression model and the order estimation criteria that will be considered. We will then apply these criteria to AM-FM signals that follow the WSF model in Sec. \ref{sec:Results_Sim}, in order to evaluate their applicability to the estimation of the number of harmonics of non stationary signals. We already mentioned the impact the WSF paradigm has made on the biomedical field. We will use our order estimation procedure on the task of denoising simulated biomedical signals, specifically pulse wave signals, in Sec. \ref{sec:Pulse}, showing the advantages of this approach over other denoising schemes. Finally, we will apply the ANH model to the task of reconstructing the ECG and respiratory signals from the Fantasia Database in Sec. \ref{sec:Fantasia}. These results demonstrate that the automatic order estimation procedure can estimate the WSF of real physiological signals contaminated with noise and that the ANH model takes into account interpatient wave-shape variability.

\section{Adaptive non-harmonic (ANH) signal modeling}\label{sec:ANH}
\subsection{Model formulation}
As we discussed previously, the simplest model we can use to describe AM-FM multicomponent signals is:
\begin{equation}
x(t) = \sum_{k=1}^{K} A_k(t)\cos(2\pi \phi_k(t)), \, \text{with} \, A_k(t),\phi_k'(t) > 0\ \forall t.
\label{eq:harmonic2}
\end{equation}

Given the phase modulation of the $k$-th component we can define the \emph{instantaneous frequency} $f_k(t)$ of that component as the first derivative of the phase, $ f_k(t) = \phi_k'(t)$. It is commonly required that $A_k(t)$ and $\phi_k'(t)$ \emph{vary slowly}, meaning that:
\begin{equation*}
|A_k'(t)|/A_k(t)< \varepsilon_1, \qquad |\phi_k''(t)|< \varepsilon_2, \qquad \forall t.
\end{equation*}

When $K>1$ we also impose a \emph{separability} condition: $|\phi_k'(t) - \phi_{k-1}'(t)|>d$, for all $t$. Note that here we are also assuming that the components are sorted by their instantaneous frequencies in an increasing fashion. The frequency content of signals modeled as in \eqref{eq:harmonic2} can be represented as a 2-D image via the squared modulus of the STFT (respectively WT), called the spectrogram (respectively scalogram), where one of the axis corresponds to the time variable $t$ and the other to the frequency $f$ (respectively scale $s$). In this domain, the energy of the signal concentrates along the instantaneous frequency of the components forming \emph{ridges}, with each individual component generating one ridge. These ridges allow the estimation of both the instantaneous frequencies and amplitudes of the components by a ridge-extraction procedure \cite{meignen2012new,colominas2020fully}.

The main limitation of the AH model	is that it represents each component as an AM-FM cosine wave. Although this condition is applicable in a variety of simple cases (pure tones, linear and non-linear chirps, etc.), some oscillatory phenomena display more complicated wave morphology. Some examples of this are biomedical signals (pulse wave, electrocardiogram (ECG), respiration-related signals) and mechanical waves (seismic waves, sound). This is aggravated even more when there are many distinct components each with their own non-sinusoidal oscillatory pattern.
In light of this situation, HT Wu~\cite{wu2013instantaneous} proposed the following model, named as the \emph{adaptive non-harmonic} (ANH) model in~\cite{wu2016modeling}, which we repeat from Sec. \ref{sec:Introduction}:
\begin{equation}
x(t) = \sum_{k=1}^K A_k(t)s_k(2\pi\phi_k(t)).
\end{equation}

Here, for each component, the cosine wave (\emph{harmonic function}) is replaced by more general $2\pi$-periodic wave-shape functions $s_k(\cdot)$ (\emph{non-harmonic}). Each WSF can be characterized by its Fourier expansion. Then:
\begin{equation}
x(t) = \sum_{k=1}^K \sum_{\ell=-\infty}^{\infty} A_k(t)\hat{s}_k(\ell) e^{2\pi i\ell\phi_k(t)},
\label{eq:WSFK}
\end{equation}

\noindent where $\hat{s}_k(\cdot)$ are the Fourier expansion coefficients for the wave shape $s_k(t)$. A particular case, advantageous for some simpler signals, is to consider only one wave-shape (i.e. $K = 1$) simplifying the model to:
\begin{equation}
\label{eq:WSF_1Comp}
x(t) = A(t)s(2\pi\phi(t)),
\end{equation}

\noindent and with $\hat{s}(\ell)$ being the Fourier coefficients of the single wave-shape $s(t)$, we obtain
\begin{align}
x(t) &= A(t) \sum_{\ell = -\infty}^{\infty} \hat{s}(\ell)e^{2\pi i\ell\phi(t)} \\ &= A(t)\sum_{\ell=0}^{\infty}\alpha_{\ell}\cos(2\pi \ell\phi(t)) + \beta_\ell\sin(2\pi \ell\phi(t)),
\label{eq:WSF1}
\end{align}
\noindent where we have $\alpha_\ell = 2\Re(\hat{s}(\ell))$ and $\beta_\ell = -2\Im(\hat{s}(\ell))$. We know that the sum of a sine and a cosine function with the same argument can be rewritten as a phase-shifted cosine, so the model becomes:
\begin{equation}
x(t) = A(t)\sum_{\ell=0}^{\infty} a_\ell\cos(2\pi \ell\phi(t) + \varphi_\ell),
\end{equation}
\noindent with $\varphi_\ell = \tan^{-1}(\beta_\ell/\alpha_\ell)$ and $a_\ell = \sqrt{\alpha_\ell^2+\beta_\ell^2}$. We note here that this model ends up as a special case of the model presented in \eqref{eq:harmonic} where the instantaneous phase of the $\ell$-th component is $\ell$ times the phase of the first component (also known as the \emph{fundamental component}). In this context, we denote each component as a \emph{harmonic} with the fundamental harmonic being $a_1\cos(2\pi\phi(t) + \varphi_1)$.

HT Wu established a number of conditions for the model in \eqref{eq:WSF1}, mainly related to the structure of the Fourier series of $s(t)$ \cite{wu2013instantaneous}: 
\begin{align*}
(\text{C}1)\ & \forall \ell\in\mathbb{N}-\{1\},\ |\hat{s}(\ell)|\leq\delta|\hat{s}(1)| \\
(\text{C}2)\ & \sum_{\ell>r_0} |\ell\hat{s}(\ell)|\leq\theta
\end{align*}

Condition C1 means that the coefficient associated with the fundamental frequency cannot be zero. Moreover, if $\delta<1$, then the first harmonic is \emph{dominant} and the fundamental component will have the highest ridge in the TF plane. We can then estimate the instantaneous frequency by detecting the most energetic ridge in the spectrogram. Although this condition might result convenient, some commonly encountered biomedical signals, like ECG or voice, may not fulfill it.

Condition C2 tells us that the coefficients of the Fourier expansion decay sufficiently fast for all $\ell$ greater than $r_0$. This allows us to approximate the wave-shape (which, in general, can have an infinite number of harmonics) by a band-limited version with only $r_0$ harmonics, with accuracy $\theta$.
\subsection{ANH model reconstruction procedure}
\textbf{Remark.} Even though the theory of WSF was originally developed in the continuous setting, we will apply it to a discretized version of $x(t)$, and it is on this setting in which we will describe the WSF reconstruction algorithm and the different model selection criteria. The discretization is performed as follows: given a continuous-time signal $x(t)$ and a total signal duration $T$, we can define a discrete signal $\mathbf{x}(n)$ of length $N$ by sampling $x(t)$: $\mathbf{x}(n) = x(n\Delta t)$, for $0\leq n\leq N-1$, where $\Delta t = T/N = 1/f_s$ is the sampling interval and $f_s$ is known as the sampling rate. The instantaneous amplitude and phase will also become discrete: $\textbf{A}(n) = A(n\Delta t)$, $\boldsymbol{\phi}(n) = \phi(n\Delta t)$. Without any loss of generality, we will also assume that all the studied signals $\textbf{x}(n)$ are zero-mean.

Given a signal as in \eqref{eq:WSF1}, if we can somehow estimate $\textbf{A}(n)$, $\boldsymbol{\phi}(n)$ and $s$, then we can reconstruct the signal. This becomes useful for modeling and processing signals that are contaminated with noise and artifacts. The estimation of $\textbf{A}(n)$ and $\boldsymbol{\phi}(n)$ has been widely studied and different approaches have been proposed, including STFT- and CWT-based ridge extraction \cite{carmona1997characterization}, synchrosqueezing transform~\cite{daubechies2011synchrosqueezed}, multitapered synchrosqueezed transform~\cite{daubechies2016conceft} and optimization of time-frequency representation~\cite{kowalski2018convex}. In our case, we estimate the amplitude and phase from a complex version of the fundamental component of the signal, explained in steps 4 and 5 of Algo. 1. Once this is done, we have to determine the wave-shape function. To do this, we solve the following functional regression problem to obtain the Fourier expansion coefficients of the WSF:

\begin{equation}
\hat{\boldsymbol{\gamma}} = \underset{\boldsymbol{\gamma}\in \mathbb{R}^{2r}}{\arg\min}\ ||\textbf{x}(n) - \mathbf{C_r}\boldsymbol{\gamma}\||^2,
\label{eq:regression}
\end{equation}

\noindent where $\mathbf{C_r}$ is an \emph{ad hoc} dictionary matrix of the form: $\mathbf{C_r} = [\textbf{c}_1,\dots, \textbf{c}_r, \textbf{d}_1, \dots,\textbf{d}_r]$, with columns $\textbf{c}_\ell = \textbf{A}(n)\cos(2\pi\ell\boldsymbol{\phi}(n))$
and $\textbf{d}_\ell = \textbf{A}(n)\sin(2\pi\ell\boldsymbol{\phi}(n))$, for $\ell = 1,\dots,r$. We solve for the coefficients column vector
$\boldsymbol{\gamma} = [ \alpha_1,\dots,\alpha_r,\beta_1,\dots,\beta_r]^T$. This least squares problem has a solution given by the normal equations:

\begin{equation}
\hat{\boldsymbol{\gamma}} = (\mathbf{C_r}^T \mathbf{C_r})^{-1}\mathbf{C_r}^T\textbf{x}.
\label{eq:normal_eq}
\end{equation}

Finally, we can reconstruct the signal with the synthesis formula 
\begin{equation}
\textbf{x}_r(n) = \mathbf{C_r}\hat{\boldsymbol{\gamma}}.    
\end{equation}
\noindent Note that we need to fix the value of $r$ beforehand in order to solve this problem. In the next section, we will propose an automatic method to estimate the optimal value of $r$ used to form the matrix $\mathbf{C}_r$.

\section{Trigonometric regression model selection criteria for the estimation of the number of harmonics}\label{sec:Criteria}

In this section, we will study different criteria for determining the number of terms (denoted as $r_0$) in a trigonometric regression model of the form
\begin{equation}
x(t) = \mu + \sum_{r=1}^{r_0}\left(\alpha_r \cos(\lambda_r t) + \beta_r\sin(\lambda_r t)\right) + u(t).
\label{eq:trigo_cont}
\end{equation}

The term $u(t)$ represents any non-sinusoidal component of the signal, like noise or trend. Constant $\mu$ can be interpreted as the mean value of the signal if $\mathbb{E}[u(t)] = 0$. We can relate this regression model to the ANH model \eqref{eq:WSF1} by replacing the linear phase argument $\lambda_r t$ in the trigonometric terms of \eqref{eq:trigo_cont} with $2\pi r\phi(t)$ and the amplitude coefficients with $A(t)\alpha_r$ and $A(t)\beta_r$, where $\alpha_r$ and $\beta_r$ are the solution to the least squares problem in \eqref{eq:regression}. We also have $\mu = 0$ as we are considering signals with zero-mean. Given these conditions, the only unknown parameter of this model is $r_0$. 

We can apply trigonometric model selection criteria to estimate the number of terms  \cite{eubank1990curve,kavalieris1994determining}. Model selection implies that we choose the model parameters so that some function of the model (and its parameters) is optimized. The simplest selection criterion is minimizing the mean square error (MSE) between the discrete-time signal $\mathbf{x}(n)$ and the model output:
\begin{equation}
\mbox{MSE}(r)\! =\! \frac{1}{N}\sum_{n=0}^{N-1} \bigg(\mathbf{x}(n) -  \mathbf{A}(n)\sum_{\ell\leq r}a_{\ell}\cos(2\pi \ell \boldsymbol{\phi}(n) + \varphi_\ell)\bigg)^2.
\end{equation}

This criterion decreases monotonically with the value of $r$, so the chosen model will always be the one with highest $r$. In the absence of noise, the MSE will reach zero when the order of the model is the correct one ($r_0$), so that terms with $j>r_0$ vanish. In the noisy case, however, these terms are no longer zero which results in a degradation of the estimated signal. Therefore, we need a criterion that allows us to establish a trade-off between the MSE and the order of the model.

Throughout the years, many such criteria have been proposed for the goal of model selection. Wax and Kailath~\cite{wax1985detection} introduced a signal detection scheme based on the information criterion proposed by Akaike (AIC)~\cite{akaike1998information} and the minimum description length (MDL), developed by Rissanen~\cite{rissanen1978modeling}:
\begin{align*}
\mbox{AIC}(r) & = -2\hat{L}(r) + 2r\\
\mbox{MDL}(r) & = -\hat{L}(r) + \frac{1}{2}r\log(N).
\end{align*}

\begin{comment}
These methods require an estimation of the maximum log-likelihood $\hat{L}$ for the parameter $r$, and then they add a bias or penalization term that depends on the order of the model. In the presence of noise, $\hat{L}(r)$ will decreases monotonically with the model order $r$ and by adding an appropriate penalization term we obtain a criterion that reaches a minimum at $r=r_0$.
\end{comment}

These methods require an estimation of the maximum log-likelihood $\hat{L}$ for the parameter $r$, and then they add a bias or penalization term that depends on the order of the model. The log-likelihood can be estimated by building a covariance matrix $\mathbf{R}$ of $\mathbf{x}(n)$ and applying the spectral theorem to obtain an eigendecomposition of $\mathbf{R}$:

\begin{equation}
\hat{L}(r) = \log\left(\frac{\prod_{i=r+1}^{N}\lambda_i^{1/(N-r)}}{\frac{1}{N-r}\sum_{i=r+1}^{N}\lambda_i}\right)^{(N-r)N},
\end{equation}

\noindent where $\lambda_1\geq\lambda_2\geq\lambda_3\geq\dots\geq\lambda_N$ are the eigenvalues of $\mathbf{R}$. Note that the term inside the parentheses is the ratio of the geometric mean of the smallest $N-r$ eigenvalues. If the signal $\mathbf{x}(n) = \mathbf{A}(n)s(2\pi\boldsymbol{\phi}(n))$ is composed of an AM-FM waveshape with $r_0$ harmonics, then the $N-r_0$ smallest eigenvalues of $\mathbf{R}$ will be equal to zero. Then $\hat{L}$ will reach a minimum at $r=r_0$ and stay constant for $r>r_0$. If the signal is contaminated with zero-mean noise, the $N-r_0$ smallest eigenvalues will now be all equal to the noise variance $\sigma^2$. In this case, $\hat{L}(r)$ will decrease monotonically with $r$ but the sharpest drop will still happen at $r=r_0$. Then, by adding an appropriate bias term we can form a criterion that reaches a minimum at $r=r_0$. Wax and Kailath mention that the number of component signals can be consistently estimated using any criteria of the form $-\log\hat{L}(r) + \alpha(N)r$ where $\alpha(N)\rightarrow \infty$ and $\alpha(N)/N\rightarrow 0$. It follows that AIC is not a consistent estimator whereas MDL is.

Even though these criteria do not work well when dealing with signals that have time-varying phases, the idea of a trade-off between model size and approximation error has been applied in the context of regression models of the form in \eqref{eq:trigo_cont}. In the rest of this section, we will describe the criteria considered in this work.

For all criteria to be discussed, we specify the range of values for the parameter $r$ to be the interval $1\leq r\leq r_{max}$. The upper bound $r_{max}$ is computed as: $r_{max} = \floor*{\frac{f_s/2}{f_{max}}}$, where $f_s$ is the sampling rate and $f_{max} = \underset{t}{\max}\ f(t)$ is the maximum value of the instantaneous frequency $f(t)$ for all $t$. 

\subsection{Wang criterion}
Wang~\cite{wang1993aic} proposed the following AIC-type criterion for the trigonometric regression model: 

\begin{equation}
\Phi_W(r) = \log \left(\mbox{MSE}(r)\right) + cr(\log N)/N,
\label{eq:wang}
\end{equation}

\noindent where $N$ is the length of the signal, $r$ is the order of the model and $c$ is a real positive constant. 

Wang gives an admissibility condition for the parameter $c$: $c>\frac{2\ \underset{\omega}{\max} \{\hat{f}(\omega)\}}{(2\pi)^{-1} \int_{-\pi}^{\pi} \hat{f}(\omega)d\omega}$, where $\hat{f}(\omega)$ is an estimation of the power spectrum of the noise. When dealing with white stationary noise, Quinn~\cite{quinn1989estimating} showed that a criterion of the form $\Phi_g(r) = N\log(\mbox{MSE}(r)) + 2rg(N)$ gives a consistent estimator of the model order if the following condition holds
\begin{equation}
\underset{N\rightarrow\infty}{\underline{\lim}} g(N)/\log N>1.
\end{equation}

By comparing $\Phi_g(r)$ to Eq. \eqref{eq:wang} we find $2g(N) = c\log N$ and the condition reduces to $c>2$. In this paper, we will study signals corrupted by white stationary noise with zero mean, so we will take values of $c \in \{2.1,5,8,12\}$.

\subsection{Kavalieris-Hannan criterion}

Based on the Wang criterion and their own previous work, Kavalieris and Hannan~\cite{kavalieris1994determining} introduced the following criterion for estimating the order of a trigonometric regression model: 
\begin{equation}
\Phi_K(r) = \log\{\hat{\sigma}_r^2(h)\} + (5r + h)(\log N)/N.
\end{equation}

\noindent Here, $\hat{\sigma}_r^2(h)$ is the linear prediction variance of the autoregressive model of order $h$ that is fitted to the residual  $\textbf{u}_r(n)=\textbf{x}(n)-\textbf{x}_r(n)$. The range of values for parameter $h$ is defined as  $1\leq h \leq H$. According to previous work by the authors~\cite{hannan1984multivariate}, if $\textbf{u}_r(n)$ can be reasonably modeled by a rational transfer function (ARMA model) we can set: $ H = (\log(N))^\alpha$ with $0<\alpha< \infty$. Without loss of generality, we choose $\alpha = 2$.

\subsection{Generalized Cross-Validation and Unbiased Risk}
\label{sec:Criteria_RSS}
Eubank and Speckman \cite{eubank1990curve} proposed two criteria based on penalizing the residuals square sum:

\noindent $\sum_{n=1}^{N} ({\textbf{x}(n) - \textbf{x}}_r(n))^2 =N\,\mbox{MSE}(r)$.
These criteria are the generalized cross-validation or GCV  $(\Phi_G$) and the unbiased risk ($\Phi_R$):

\begin{align}
\Phi_G(r) &= N^2 \mbox{MSE}(r)/(N-2r-1)^2,\\
\Phi_R(r) &= \mbox{MSE}(r) + 2\hat{\sigma}^2(2r + 1)/N,
\end{align}
\noindent where $\hat{\sigma}^2$ is an estimator of the noise variance. In this work, we use a variance estimator based on the STFT~\cite{donoho1994ideal,pham2018novel}:
\begin{equation}
\hat{\sigma}^2 = \Bigg(\frac{\text{median}(|\Re(\textbf{F}_\textbf{x}^\textbf{g}(n,k))|)}{0.6745\|\textbf{g}\|_2}\Bigg)^2,
\label{eq:thresh}
\end{equation}

\noindent were $\|\textbf{g}\|_2$ is the $\ell^2$ norm of the window $\textbf{g}(n)$.

In the subsequent simulations and experiments, we compute the four order selection criteria detailed in this section, using Algo. 1 to estimate the reconstructed signal $\overline{\textbf{x}}_r$, for the range $0\leq r \leq r_{max}$ and choose the optimum value $r^*$ as follows
\begin{equation}
\label{eq:opt_model}
r^* = \underset{r\ \in\ \left[1,r_{max}\right]}{\text{argmin}}\ \Phi_{X}(r);
\end{equation}

\noindent for $X \in \{G,R,W,K\}$. We summarize the reconstruction procedure in Algo. 1. We put particular emphasis on steps $6$ to $12$, where we carry out the optimal WSF order estimation procedure using trigonometric regression model selection criteria.

\textbf{Algorithm 1} WSF signal analysis and reconstruction with adaptive model order
\vspace{-2mm}

\hrulefill

\label{al:WSF}
\begin{algorithmic}[1]
	\REQUIRE Signal $\textbf{x}(n)$. STFT analysis window $\textbf{g}(n)$. Number of frequency bins $K$. Ridge extraction parameters $\lambda$, $\mu$ and maximum frequency bin jump $I$. Reconstruction bandwidth $\Delta$. Maximum admissible order $r_{max}$. Model order estimation criteria $X$ for $X\in \{G,R,W,K\}$. %ponemos todas las entradas: señal, parámetros, etc.
	\STATE Given a signal $\textbf{x}(n)$ of the form in \eqref{eq:WSF1} compute its short-time Fourier transform (STFT) by
	\begin{equation*}
	\textbf{F}^\textbf{g}_\textbf{x}(n,k) = \sum_{\nu = 0}^{N-1} \textbf{x}(\nu)\textbf{g}(\nu-n)e^{-2\pi i\frac{k}{K}\frac{N}{T}(\nu-n)},
	\end{equation*}
	with $k = 0,1,\dots,K-1$ and $n = 0,1,\dots,N-1$.
	\STATE Compute the de-shape STFT $\mathbf{W}_\textbf{x}(n,k)$. In this representation, the harmonics of the WSFs are suppressed leaving only the fundamental component. For more information, see \cite{lin2018wave} and the Supplemental Material.
	\STATE Given the T-F representation $\mathbf{W}_\textbf{x}$, find the ridge associated with the fundamental frequency $\textbf{c}(n) \approx \boldsymbol{\phi}'(n)$. To do this, solve the following functional problem: 	
	\begin{align*}
	\textbf{c}^*&(n) = \underset{\textbf{c}(n)}{\arg \max }\Bigg(\sum_{n=0}^{N-1} |\textbf{W}_\textbf{x}(n,\textbf{c}(n))|^2  -\sum_{n=1}^{N-1} \lambda | \textbf{c}(n) - \textbf{c}(n-1)|^2
	 -\sum_{n=2}^{N-1}\mu |\textbf{c}(n) - 2\textbf{c}(n-1) + \textbf{c}(n-2)|^2 \Bigg).
	\end{align*}
	\noindent We find an approximated solution by implementing a greedy ridge detection procedure proposed by Meignen \emph{et al.}~\cite{meignen2012new} to obtain an approximation of $\textbf{c}^*(n)$. In this, the maximization is limited to a frequency interval $I_n = \{c(n-1)-I,\dots,c(n-1)+I\}$ around the ridge estimation at the previous time step. $I$ defines the maximum frequency bin jump for consecutive steps. 
	\STATE Reconstruct the fundamental component of $\textbf{x}(n)$ by integrating the STFT around the ridge $\textbf{c}^*(n)$:
	\begin{equation*}
	\textbf{y}(n) = \frac{1}{\textbf{g}(0)}\sum_{|k-\textbf{c}^*(n)|<\Delta} \textbf{F}^\textbf{g}_\textbf{x}(n,k)
	\end{equation*}
	\STATE Compute the estimates for the amplitude $\tilde{\textbf{A}}(n) = |\textbf{y}(n)|$ and phase modulation $\tilde{\boldsymbol{\phi}}(n)=\phase{\textbf{y}(n)}$.
	\FOR{$r=1$ to $r_{max}$}
	\STATE Construct $\mathbf{C}_r$ as described in section $\ref{sec:ANH}$.
	\STATE Compute $\hat{\boldsymbol{\gamma}}$ using Eq. \eqref{eq:normal_eq}
	\STATE Compute $\mathbf{x}_r$, the $r$-order approximation of $\mathbf{x}$.
	\STATE Compute the order selection criteria $\Phi_X(r)$.
	\ENDFOR
	\STATE Find the optimum model order $r^*$ using Eq. \eqref{eq:opt_model}.
	\STATE Construct $\mathbf{C_{r^*}}$ as described above, solve for $\hat{\boldsymbol{\gamma}}_{r^*}$ using Eq. \eqref{eq:normal_eq} and then compute $\textbf{x}_{r^*}(n) = \mathbf{C_r}\hat{\boldsymbol{\gamma}}_{r^*}$.
	\ENSURE	Reconstructed signal $\textbf{x}_{r^*}(n)$ and Fourier coefficients vector $\hat{\boldsymbol{\gamma}}_{r^*}$.
\end{algorithmic}

\vspace{-3mm}
\hrulefill

\section{Results on simulated AM-FM signals}
\label{sec:Results_Sim}
The criteria we described on the previous section were developed for model \eqref{eq:trigo_cont}, which considers \emph{non-modulated} signals. We will illustrate here that the same criteria can be applied to AM-FM signals, such as those modeled with \eqref{eq:WSF1}.

In all cases, signals were discretized by sampling them at a rate of $f_s = 3000$ samples per second. We computed the STFT as shown in Algo. 1 using a discrete Gaussian window $\textbf{g}(n) = e^{-2\sigma n^2}$ with $\sigma = 10^{-4}$ and discrete frequency interval $0\leq k \leq f_s/2$ with frequency step of $f_s/N$. In the ridge extraction algorithm, we considered $\lambda = \mu = 0.1$ and set the maximum frequency bin jump at $I = 10$.For the amplitude and phase estimation procedure based on integrating around the fundamental ridge, we used $\Delta = 50$ frequency bins.

We note here that since we know the exact values of $A(t)$ and $\phi(t)$ of our simulated signal, we could use them to reconstruct the signal instead of estimating them  using the ridge extraction procedure. The presence of noise causes error in the estimations, which degrades the quality of the reconstructed signals. Nevertheless, we chose to estimate $A(t)$ and $\phi(t)$ using Algo. 1, in a more realistic setting, in order to evaluate the reconstruction procedure as a whole. 

\subsection{Analysis of model selection criteria}\label{sec:Analysis1D}

\begin{figure}
	\centering   
	\includegraphics[width=0.5\textwidth]{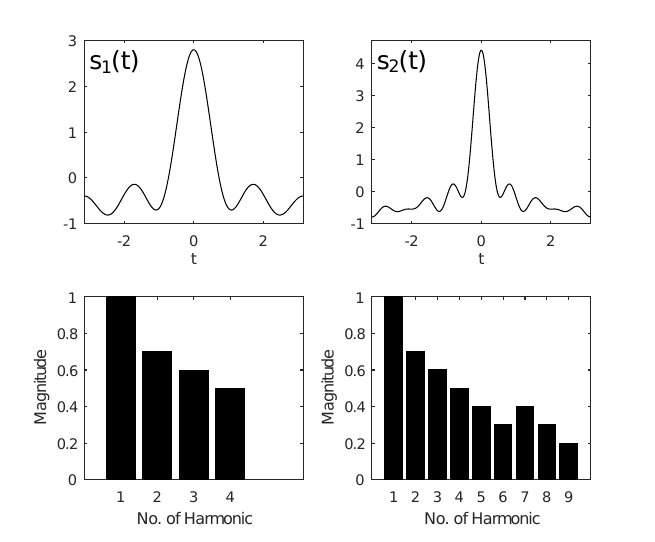}
	\includegraphics[width=0.5\textwidth]{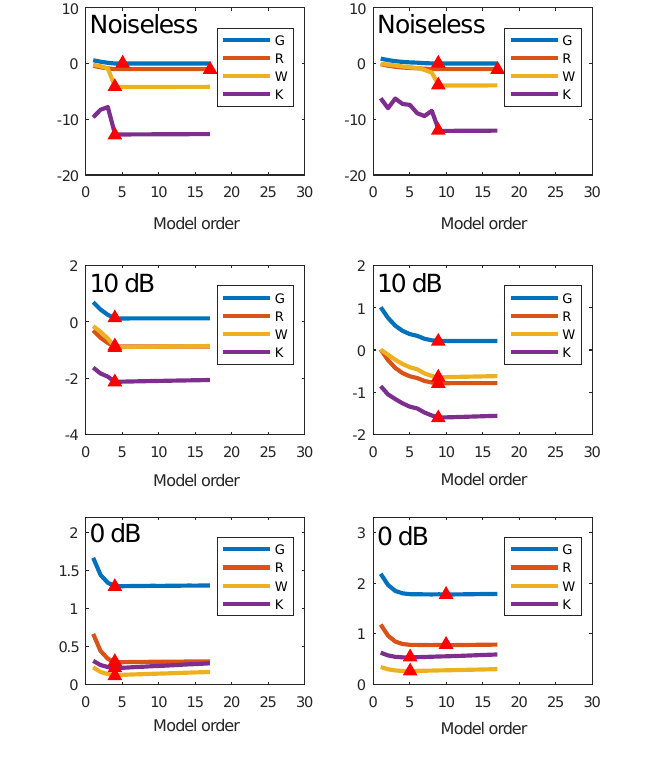}
	\caption{Top Row: Wave-shapes of $x(t)$ with $r_0 = 4$ (left) and $r_0 = 9$ (right). Second Row: Fourier coefficients of the corresponding wave-shape. Third Row: Results of the order estimation of the wave shapes in the absence of noise. Fourth Row: Same as above but with SNR $= 10$ dB. Fifth Row: Results for SNR = $0 $ dB. In all cases, for $\Phi_W(r)$ and $\Phi_K(r)$, the curve shown is obtained by fixing the value of $c$ or $h$, respectively, to the one where the global optimum of the criterion is reached. References: G - Generalized Cross-Validation. R - Unbiased Risk. W - Wang Criterion. K - Kavalieris-Hannan Criterion.}
	\label{fig:Exp1}
\end{figure}

We started by analyzing an amplitude-modulated signal with nonlinear frequency modulation and non-sinusoidal oscillation pattern:
\begin{equation}\label{eq:first_signal}
x(t) = \left(1+0.05\sqrt{t}\right)s\left(2\pi\left(70t + \frac{15}{2\pi}\cos(2\pi t)\right)\right) + n(t),
\end{equation}

\noindent with $t\in[0,1]$, and where the term $n(t)$ is zero-mean white Gaussian noise. The fundamental instantaneous frequency is: $\phi'(t) = 70 - 15\sin(2\pi t)$. We considered, in this first example, two band-limited wave-shape functions, i.e. $s$ such that $\hat{s}(\ell) = 0$ for $\ell>r_0$. The corresponding wave-shapes and their Fourier coefficients (with $r_0 = 4$ and $r_0 = 9$ respectively) are shown in top two rows of Fig. \ref{fig:Exp1}. The amount of noise is characterized by the signal-to-noise ratio (SNR) in decibels (dB). We considered two levels of noise, $0$ dB and $10$ dB. As a reference, we also studied the noise-free case.

In Fig. \ref{fig:Exp1} we show the result of computing the model selection criteria for a realization of the signal $x(t)$ from Eq. \eqref{eq:first_signal}. First, we look at the results for the noiseless case in the third row. We observe that, for the two analyzed signals, both $\Phi_W(r)$ and $\Phi_K(r)$ display a sharp drop at $r=r_0$ and the minimum is reached for this order. $\Phi_G(r)$ has a smoother transition and the minimum is reached at $r^*= 5$ and $r^*= r_0 = 9$, for $s_1(t)$ and $s_2(t)$ respectively. As the penalization term of $\Phi_R(r)$ is dependent on the estimated noise variance and we have $\hat{\sigma}^2 = 0$ in this case, then $\Phi_R(r) = \mbox{MSE}(r)$ so the criterion decreases monotonically with $r$.

When the SNR is moderate ($10$ dB), as shown in the fourth row of Fig. \ref{fig:Exp1}, the sharp drop of $\Phi_W(r)$ and $\Phi_K(r)$ is maintained only for the case of $r_0=4$. For both wave-shape functions, all criteria estimate the true order correctly.

Finally, we studied the case contaminated with a high level of noise (SNR of $0$ dB), shown in the last row of Fig. \ref{fig:Exp1}. Under this condition, the task of detecting the number of harmonics in the trigonometric model becomes more difficult. In the case of $s_1(t)$ ($r_0 = 4$), all criteria are able to recover the correct model order. For $s_2(t)$ (for which we have $r_0=9$), we start seeing underestimation for $\Phi_W(r)$ and $\Phi_K(r)$, with $r^* = 4$ for both. In contrast, $\Phi_G(r)$ and $\Phi_R(r)$ show an slight overestimation ($r^* = 10$).  
Based on this results, we can see that all criteria are able to correctly estimate the model order $r_0$ under the time-varying phase assumption, although some difficulties might arise when both the level of noise and the number of harmonics are high.

\subsection{Wave-shape estimation for AM-FM signals}

The simulations described in the previous section correspond to a single realization of a noisy signal with a fixed number of harmonic components. The main causes of variation in the model selection procedure are noise and the relative amplitude of the harmonics, mainly the high-frequency ones. The criteria may underestimate the order if the relative amplitude of the higher-order harmonics is small when compared to the noise level. In order to minimize the effect of these factors, we generated a large number of realizations and randomized the amplitude of the second to last harmonics in the range $[0.1,0.9]$ while maintaining the fundamental harmonic as dominant (i.e. $\hat{s}(1) = 1$). In that way, we synthesized different wave-shape functions $s$, constructing signals of the form $x(t)=A(t)s(2\pi\phi(t))$, with $A(t) = 1+0.05\sqrt{t}$. We also considered different types of frequency modulation:

\begin{itemize}
	\item $\phi_{NM}(t) = 100t$ (no modulation).
	\item $\phi_{LM}(t) = \frac{15}{2}t^2 + 50t$ (linear modulation).
	\item $\phi_{SM}(t) = -\frac{15}{2\pi}\cos(2\pi t) + 70t$ (sinusoidal modulation).
\end{itemize}

\begin{figure*}
	\centering
	\includegraphics[width=0.9\textwidth]{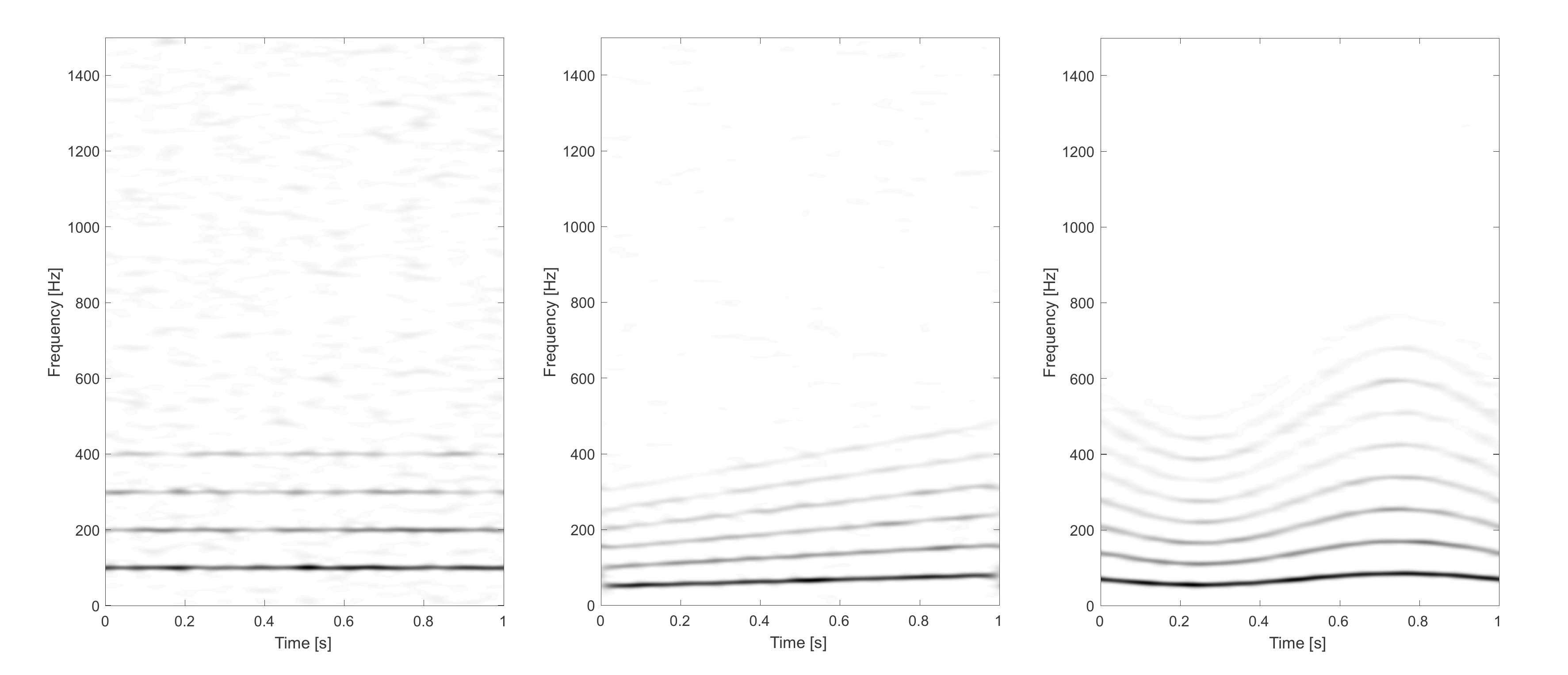}
	\caption {Left: Spectrogram of $x_{NM}(t)$ contaminated with $0$ dB additive noise and $r_0 = 4$. Middle: Spectrogram of $x_{LM}(t)$ contaminated with $5$ dB additive noise and $r_0 = 6$. Right: Spectrogram of $x_{SM}(t)$ contaminated with $10$ dB additive noise and $r_0 = 9$.}
	\label{fig:PhaseTF}
\end{figure*}

\begin{figure*}
	\begin{center}
		\includegraphics[width=0.8\textwidth]{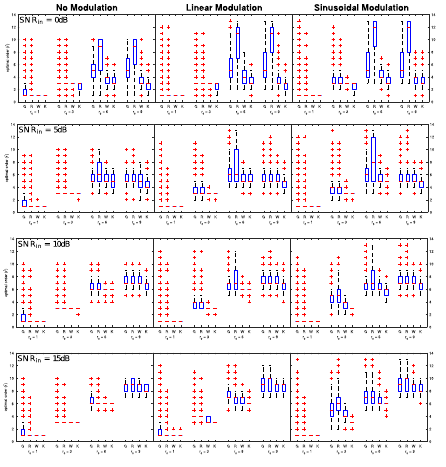}
		\caption{Left: results of the second experiment for non-modulated signals $x_{NM}(t)$. Middle: results for $x_{LM}(t)$. Right: results for $x_{SM}(t)$. In each column, the box width indicates the variance of that set of realizations. The outliers in each condition set are represented with red crosses. The red line inside each box denotes the median value of the box. Abbreviations: G - Generalized Cross-Validation. R - Unbiased Risk. W - Wang criterion. K - Kavalieris-Hannan criterion.}
		\label{fig:AllMod}
	\end{center}
\end{figure*}

We considered two additional SNRs: $5$ dB and $15$ dB. We set the possible values of $r_0$ to $1$, $3$, $6$ and $9$. We used the same parameters for the STFT and ridge extraction procedure as in the previous section. In total, $1000$ realizations for each combination of SNR, $r_0$ and $\phi(t)$ were generated and all criteria computed for every case. We show the spectograms for noisy signals of each type of phase modulation in Fig. \ref{fig:PhaseTF} using different number of harmonics. It is clear that as the noise level increases, the higher order harmonics that form lower ridges in the T-F are obscured by the noise, making their detection more challenging. The results are presented in Fig. \ref{fig:AllMod} through the use of boxplots.

On the left column of Fig. \ref{fig:AllMod} we see the results for $x_{NM}(t) = A(t)s(2\pi\phi_{NM}(t))$. When the SNR is low ($0$ dB), $\Phi_G(r)$ and $\Phi_R(r)$ perform well as order estimators for $r_0\leq 6$, presenting only a few outliers. $\Phi_W(r)$ and $\Phi_K(r)$ have comparable results in this range, but they start showing a bias towards lower model orders. When the signal is composed of a large number of harmonic components ($r_0=9$), all criteria underestimate the model order. This condition is worse for both $\Phi_W(r)$ and $\Phi_K(r)$. As the SNR improves, this bias towards lower model orders diminishes. For the $5$ dB case, both $\Phi_G(r)$ and $\Phi_R(r)$ estimate the correct order when $r_0 = 9$. At higher SNRs ($10$ and $15$ dB) all criteria estimate the model order accurately, with only $\Phi_K(r)$ presenting a slight underestimation at $10$ dB.

Next, we show the results for $x_{LM}(t) = A(t)s(2\pi\phi_{LM}(t))$ in the middle column of Fig. \ref{fig:AllMod}. Here we start seeing the effect that the phase modulation has on the model order estimation. For lower SNRs ($0$ dB and $5$ dB) the results are comparable to the non-modulation case. When we compare the graphs (always comparing to the non-modulation case) for $10$ dB and $15$ dB, we see more variance (represented by the increased box width). Nevertheless, the median values for these boxes are close to $r_0$. For these SNR levels, $\Phi_G(r)$ and $\Phi_R(r)$ start showing a tendency for slight overestimation of the model order when $r_0$ is low ($r_0 = 3$).

Finally, the results for $x_{SM}(t) = A(t)s(2\pi\phi_{SM}(t))$ are presented on the right column of Fig. \ref{fig:AllMod}. We can see an increased variance of the results across the board. At lower SNRs, we see a similar behavior to the previous cases, with $\Phi_W(r)$ and $\Phi_K(r)$ noticeably underestimating the model order for $r_0 = 9$. As the SNR improves, the estimates get closer to the correct value. We can observe a tendency of overestimation for $\Phi_G(r)$ and $\Phi_R(r)$ when $r_0 = 3$, regardless of noise level.

In general, all criteria give estimates that are close to $r_0$ in most cases. Under the most favorable condition (SNR of $15$ dB) the estimates are within the range of $r_0\pm 1$ harmonic. When the noise is high, the best estimates are obtained when the numbers of relevant harmonics is low. We also note that $\Phi_G(r)$ and $\Phi_R(r)$ are more prone to overestimation, whereas $\Phi_K(r)$ has a higher chance of underestimating $r_0$.

\subsection{Model order estimation in multicomponent signals}
In these experiments, we extended the automatic estimation of the number of harmonic components in the ANH model to the multicomponent case, that is, when $K>1$ in \eqref{eq:WSFK}. Particularly, we considered the superimposition of two synthetic non-sinusoidal signals with frequency modulation. In this case, we needed to determine two unknown parameters: $r_1$ and $r_2$, the number of harmonics for the first and second WSF, respectively. To do this, we modified the model selection criteria by replacing the scalar model order variable $r$ with a two-dimensional variable $(r_1,r_2)$. The definition of the modified ``bidimensional'' criteria are the following:
\begin{align*}
\Phi_W(r_1,r_2) & = \log\left( \mbox{MSE}(r_1,r_2)\right) + c(r_1+r_2)(\log N)/N,\\
\Phi_K(r_1,r_2) & = \log\{\hat{\sigma}_{r_1,r_2}^2(h)\} + (5(r_1+r_2) + h)(\log N)/N,\\
\Phi_G(r_1,r_2) & = N^2 \mbox{MSE}(r_1,r_2)/(N-2(r_1+r_2)-1)^2,\\
\Phi_R(r_1,r_2) & = \mbox{MSE}(r_1,r_2) + 2\hat{\sigma}^2(2(r_1+r_2) + 1)/N,
\end{align*}

\noindent where the residuals were computed using \eqref{eq:WSFK} by adding the first $r_1$ terms of the first component and the first $r_2$ terms of the second component. The estimation of the Fourier coefficients was done in the same way as before, only now the matrix $\mathbf{C_{r_1,r_2}}$ included the two instantaneous amplitudes and the two instantaneous phases.
The criteria were then computed for each duple $(r_1,r_2)$ in the rectangular domain $1\leq r_1\leq r_{1,max}, 1\leq r_2\leq r_{2,max}$ and we chose the minimum coordinates $(r_1^*,r_2^*)$ for each criterion.

\begin{figure}
	\centering
	\includegraphics[width=0.7\textwidth]{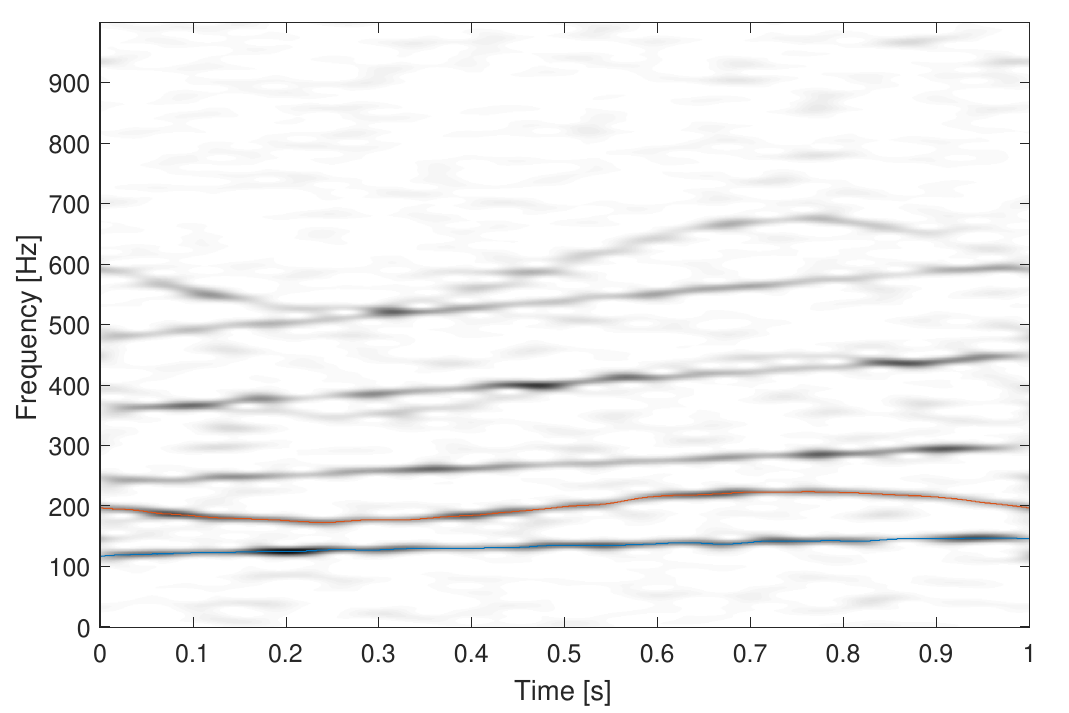}
	\caption{Spectrogram of the multicomponent signal $s(t) = s_1(t) + s_2(t)$ contaminated with $0$ dB additive noise. The fundamental ridges of components $s_1(t)$ and $s_2(t)$ are highlighted in blue and red, respectively.}
	\label{fig:Spectro_Multi}
\end{figure}

In these simulations, we generated a superimposition of chirps $s(t) = s_1(t) + s_2(t)$, with 
\begin{equation}
s_1(t) = \sum_{\ell=1}^{r_1} a_{1,\ell}\cos(2\pi\ell(120t +15t^2))   
\end{equation}

\noindent being a sum of linear chirps, and 
\begin{equation}
s_2(t) = \sum_{\ell = 1}^{r_2} a_{2,\ell}\cos\left(2\pi\ell \left(200t+\dfrac{25}{2\pi}\cos(2\pi t)\right)\right) 
\end{equation}

\noindent being a sum of sinusoidal chirps. These signals were discretized at a sampling rate of $3000$ Hz in the interval $[0,1]$. We fixed $r_1 = 4$ and $r_2 = 3$, set $a_{1,1} = a_{2,1} = 1$, and randomized the amplitude of the second to last harmonic coefficients for each realization into the interval $[0.1,0.9]$ (by doing so, we kept the fundamental harmonic dominant). The same parameters from section \ref{sec:Analysis1D} were used to compute the STFT, extract the fundamental ridges and reconstruct the signal. We set $c = 2.1$ for $\Phi_W(r_1,r_2)$ and $h = (\log N)^2$ for $\Phi_K(r_1,r_2)$.  We added noise to $s(t)$ at 3 different SNR levels: $0$, $10$ and $20$ dB. We computed the criteria for $1000$ realizations of noisy signals for each noise level and choose the optimal order pair $(r_1^*,r_2^*)$. We show the spectrogram of a realization of this multicomponent signal with SNR of $0$ dB in Fig. \ref{fig:Spectro_Multi}. It becomes clear that higher order harmonics of both components overlap in the time-frequency plane for this type of signals.

\begin{figure*}[t]
	\centering
	\includegraphics[width=0.9\textwidth]{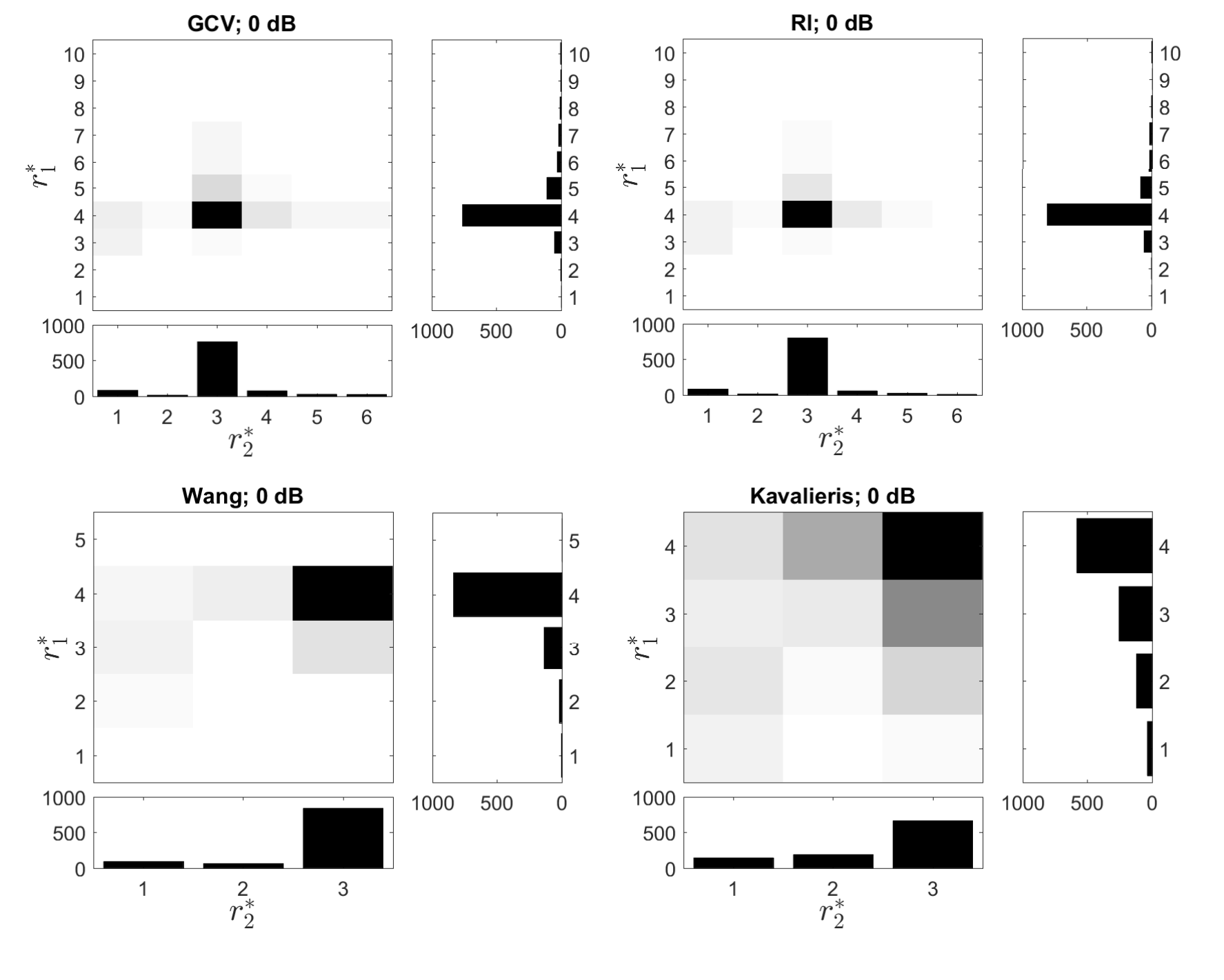}
	\caption{Results for the ANH model estimation for the multicomponent signal $s(t) = s_1(t) + s_2(t)$ with SNR of $0 $ dB. For each criterion, the intensity map and the marginal histograms are shown. All criteria estimate the order for both components correctly as shown by the modes in the histograms.}
	\label{fig:Results2Sigs2}
\end{figure*}

The results of these simulations with SNR of $0$ dB are presented in Fig. \ref{fig:Results2Sigs2}. We show the 2D intensity map alongside the marginal histograms for $r_1^*$ and $r_2^*$, where we can see that the criteria provide good estimators for the model order parameters $r_1$ and $r_2$, as the modes of the histograms are the correct model orders for all criteria. $\Phi_G(r_1,r_2)$ (denoted as GCV) and $\Phi_R(r_1,r_2)$ (RI) have a wider spread and a greater tendency to overestimate the model order than $\Phi_W(r_1,r_2)$ (Wang) and $\Phi_K(r_1,r_2)$ (Kavalieris). The results for higher SNRs are not shown (for space reasons) but they improve as the SNR increases, being more concentrated at the correct order pair. These results indicate that the trigonometric regression model selection criteria are well-suited for the automatic detection of the different WSFs that compose a signal with time-varying instantaneous frequencies.

\section{Denoising of simulated pulse signals}\label{sec:Pulse}
\begin{figure*}
	\centering
	\includegraphics[width=0.7\textwidth]{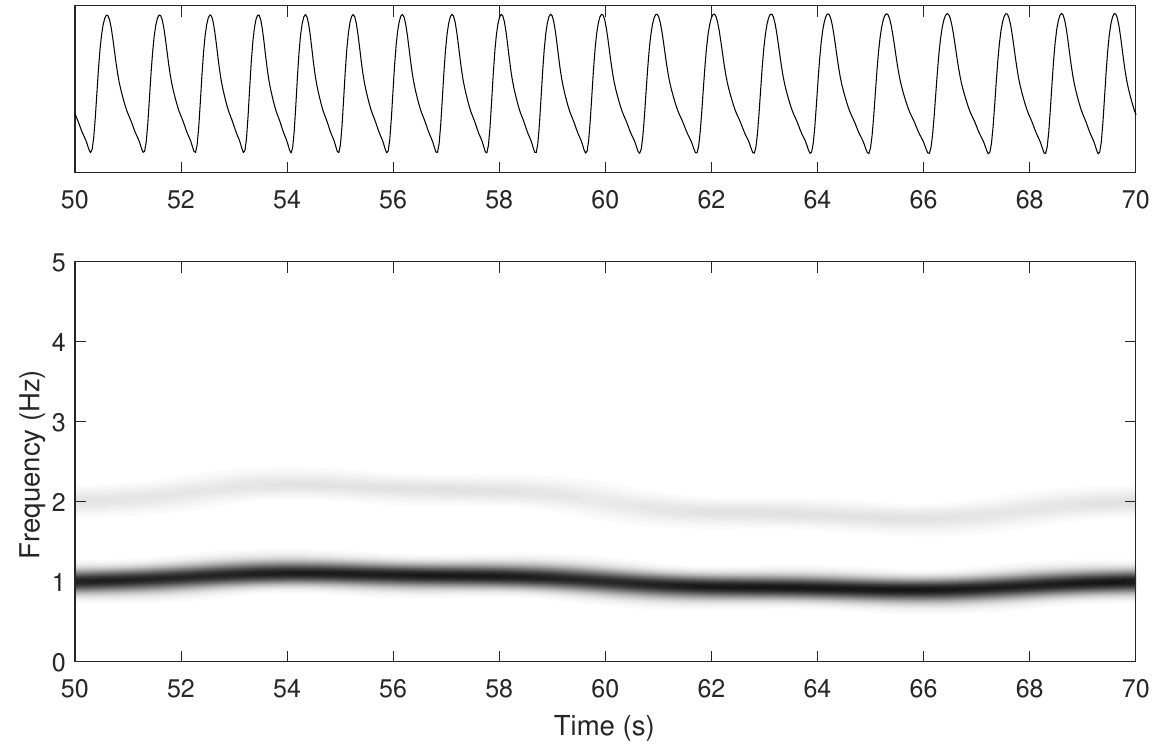}
	\caption{Top: 20-second segment of a pulse wave signal $x(t)$ given by Eq. \eqref{eq:waveshapepulse}. This is one of the possible oscillatory pattern corresponding to the abdominal aorta measurement site. Bottom: Spectrogram of $x(t)$, the ridges corresponding to the fundamental component $\phi'(t)$ and the second harmonic $2\phi'(t)$ are visible.}
	\label{fig:pulsewave}
\end{figure*}
Time-frequency techniques have been applied to numerous signal processing tasks in the biomedical field, including signal modeling~\cite{wu2016modeling}, component separation \cite{su2017extract} and denoising \cite{pham2018novel}. In this section, we will evaluate the viability of the trigonometric order selection criteria as a way to determine the optimal number of harmonics for the wave-shape $s$ in model \eqref{eq:WSF1}, when applied to biomedical signals contaminated with noise. Particularly, we will analyze signals from the Pulsewave Database\footnote{https://zenodo.org/record/3275625\#.YSk9xt-vHIU} (PWDB)~\cite{pwdb2019,charlton2019modeling}. This database collects pulse wave signals generated from a model that simulates the haemodynamics of healthy individuals. With this model, pulse waves corresponding to different sections of the circulatory system can be generated. For the purposes of these simulations we will focus on pulse waves from the abdominal aorta region.

\subsection{Experiment design}

In order to evaluate the performance of the ANH model when dealing with biomedical signals contaminated with noise, we synthesized pulse wave recordings based on the waveforms contained in the PWDB and added a known amount of noise to these signals.Then, we applied the ridge extraction procedure and reconstructed the signals using the ANH model automatically estimating $r_0$, resulting in the reconstructed versions $x_r(t)$. Finally, we computed the output SNR as follows: $\text{SNR}_{out} = 20\log\left(\frac{\|x(t)\|_2}{\|x(t)-x_{r}(t)\|_2}\right)$, 
where $\|\cdot\|_2$ is the $L^2$ norm. We also compared the results of applying this procedure to the denoising scheme based on thresholding the STFT, using a robust threshold \cite{pham2018novel}. For our simulations, we applied both soft and hard thresholding, with the threshold set at $ \eta = 3\hat{\sigma}\|g\|_2$, with $\hat{\sigma}$ given by Eq. \eqref{eq:thresh} \cite{pham2018novel}.

The synthesized pulse wave signals have the  form
\begin{equation}
x(t) = (1+0.02\sqrt{t})s(2\pi\phi(t)), 
\label{eq:waveshapepulse}
\end{equation}

\noindent with $s$ being a pulse wave-shape randomly picked from the database, and $\phi(t)$ a synthetic phase function  that emulates the phase associated with physiologic heart rate variability (HRV):
$\phi'(t) = 1 + 0.035a\sin(2\pi f_{L} t) + 0.035b\sin(2\pi f_{H} t),$
\begin{figure*}[t]
	\centering
	\includegraphics[trim = 50 50 0 0, clip, width=0.75\textwidth]{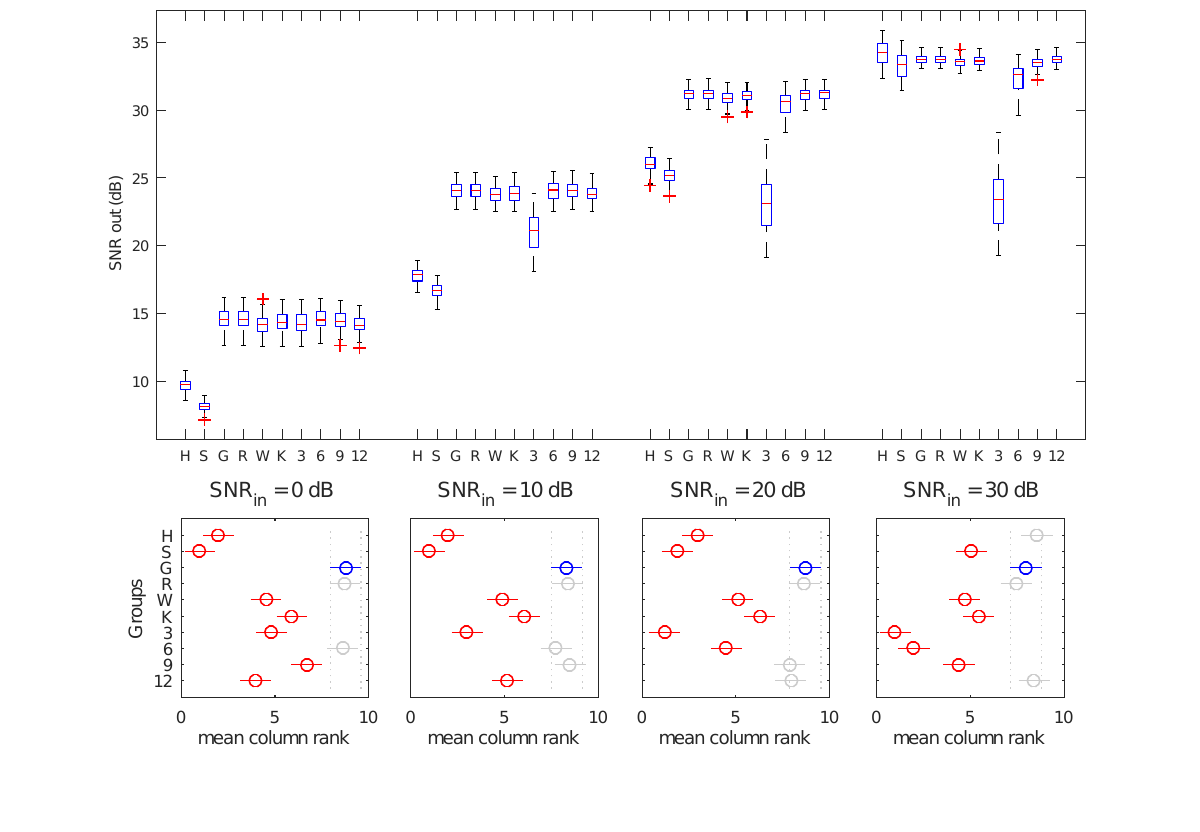}
	\caption{Results for the denoising of the pulse wave signal for each reconstruction method and each SNR$_{in}$ level. Top: Box plots showing SNR$_{out}$ as a function of SNR$_{in}$ for all the methods. Bottom: Mean column ranks obtained for the Friedman test with the corresponding confidence intervals. In each box, the mean rank for the Generalized Cross-Validation criterion is shown in blue. The methods that have a significant difference in mean rank with respect to GCV are in red.  Abbreviations: H - Hard-thresholding. S - Soft-thresholding. G - Generalized Cross-Validation. R - Unbiased Risk. W - Wang criterion. K - Kavalieris-Hannan criterion. 3 - fixed order $r = 3$ (same for 6, 9 and 12).}
	\label{fig:boxpulse}
\end{figure*}
\noindent with $f_{L}$ corresponding to the frequency associated with the low-frequency ($0.01-0.15 Hz$) energy content of the HRV, and $f_{H}$ associated with the high-frequency ($0.15-0.4Hz$) energy content. For this experiment, we fixed $f_{L} = 0.05$ Hz  and $f_{H} = 0.2$ Hz. The coefficients $a$ and $b$ were chosen so that the low-frequency to high-frequency power ratio (LHPR) is $3$ and the maximum deviation from the mean frequency ($\overline{f} = 1$ Hz) is $\pm 0.035 Hz$, in order to simulate the HRV of a healthy individual as presented in~\cite{kamen1996poincare}. We consider a signal duration of $T = 120$ seconds, sampled at $f_s = 58.33$ Hz for a total signal length of $7000$ samples. The STFT is computed as described before, with the standard deviation of the window function chosen adaptively using R\'enyi entropy method \cite{Baraniuk2001,meignen2020use}. The bandwidth for the reconstruction formula was set at $\Delta = 50$ frequency bins. The parameters chosen for ridge extraction where the same as in Sec. \ref{sec:Results_Sim}.

For these simulations: i) We randomly sampled $70$ pulse wave-shapes $s_{\{i:1\leq i\leq 70\}}$ from the abdominal aorta section of the PWDB; ii) For each wave-shape, we generated an AM-FM signal $x_i(t)$ using \eqref{eq:waveshapepulse} (we show an example in Fig. \ref{fig:pulsewave}) and then added zero-mean Gaussian noise at 4 different input SNR levels: $0$, $10$, $20$ and $30$ dB; iii) For each noisy signal, we computed the criteria described in Sec. \ref{sec:Criteria} and applied the reconstruction algorithm from Sec. \ref{sec:ANH}, using the minimum of each criterion as the optimal order. These reconstructed signals are denoted as $x_{r,i}(t)$; iv) As a reference, we also reconstruct versions of $x_i(t)$ with a fixed number of harmonics, considering the following values: $r = 3$, $6$, $9$ and $12$; v) We compared our reconstruction technique to the denoising based on STFT-thresholding, we set a threshold $\eta$ and reconstructed the signal by numerical integration over frequency of the thresholded STFT $\tilde{\mathbf{F}}$: $\mathbf{x}_{HT,i}(n) = \frac{2}{g(0)}  \Re(\sum_{k=1}^K \tilde{\mathbf{F}}(n,k))$. We also reconstruct the signal the same way using a soft threshold; vi) we computed $\text{SNR}_{out}$ for all reconstructed signals.

\subsection{Results for the denoising task}
The results of these simulations are presented in Fig. \ref{fig:boxpulse}. We can see that at every noise level there is a considerable gain in SNR when applying the adaptive order estimation algorithms. We also note that, at a given input SNR, the results are very similar for all the adaptive model order estimation methods.

Regarding fixed order reconstructions, we see that they have comparable results for low SNR ($0$ dB). As the SNR improves, the lower order models have poorer performance. The reason for this is that, as the amount of noise decreases, low amplitude harmonics can be better detected in the time-varying spectral content of the signal. When the noise is high, including more harmonics in the reconstruction will results in more noise being preserved, which degrades the output SNR. 

In the $0$ dB portion of Fig. \ref{fig:boxpulse} we see that the median for fixed orders $r = 9$ and $r = 12$ is lower than for $r = 6$.  When there is little noise (increasing SNR), it is easy to see that higher order models will provide a better approximation to the original signal. This becomes evident for SNRs of $20$ and $30$ dB. At $30$ dB the median values of $r = 12$ and $r = 9$ are considerably higher than those of $r = 6$ and $r = 3$.

Regarding the thresholding techniques, we can see that the hard threshold gives better results than the soft threshold at all SNR levels. In addition, both hard and soft thresholding have poorer performance than the rest of the methods except for the highest SNR ($30$ dB), where they give comparable results to $\Phi_G(r)$ and $\Phi_R(r)$, and fixed order reconstruction with $r = 12$. 
\subsection{Statistical analysis of the differences between methods}
In order to evaluate the statistical significance in the differences between the considered criteria, we applied a Friedman test for nonparametric analysis of variance (ANOVA) to the data from the top panel of Fig. \ref{fig:boxpulse}. We considered a combined significance level $\alpha_H = 0.05$ and used Bonferroni correction to account for the multiple comparisons, obtaining a significance level of $\alpha_B = 0.05/45 = 0.0011$. The resulting mean ranks and their confidence intervals for all considered criteria are shown in the bottom row of Fig. \ref{fig:boxpulse}. Across all SNR levels, both $\Phi_G(r)$ (noted as G) and $\Phi_R(r)$ (R) are consistently ranked higher than the other adaptive trigonometric model order estimation criteria. We also see that as the SNR improves, the best fixed order changes. This indicates that the \emph{correct} model order depends on both the wave-shape and the amount of noise, when dealing with noisy signals. When the noise is high, models with lower order have better results. Conversely, higher order models tend to perform better when the signals are less noisy. Both $\Phi_G(r)$ and $\Phi_R(r)$ are able to adaptively choose the best model order independently of the noise level and have the best (or tied for best) results when compared to the other criteria, fixed order models and the thresholding methods in each case.

\section{Adaptive Wave-shape Estimation in Real-world Physiological Signals}\label{sec:Fantasia}

In the previous section, we showed that the ANH model can be used to denoise simulated biomedical signals, and illustrated the value of objectively selecting the parameter $r$. In this section, we shift the focus to the task of modelling real physiological signals. We will show that it is necessary to adaptively choose the order $r$ of the ANH model for each signal in order to account for interpatient wave-shape variability. We will analyze physiological signals from the Fantasia Database\footnote{https://physionet.org/content/fantasia/1.0.0/}~\cite{fantasia2003}, namely ECG and respiratory signals, available through the research resource Physionet~\cite{goldberger2000physiobank}.

\begin{figure}
	\centering
	\includegraphics[width=0.45\textwidth]{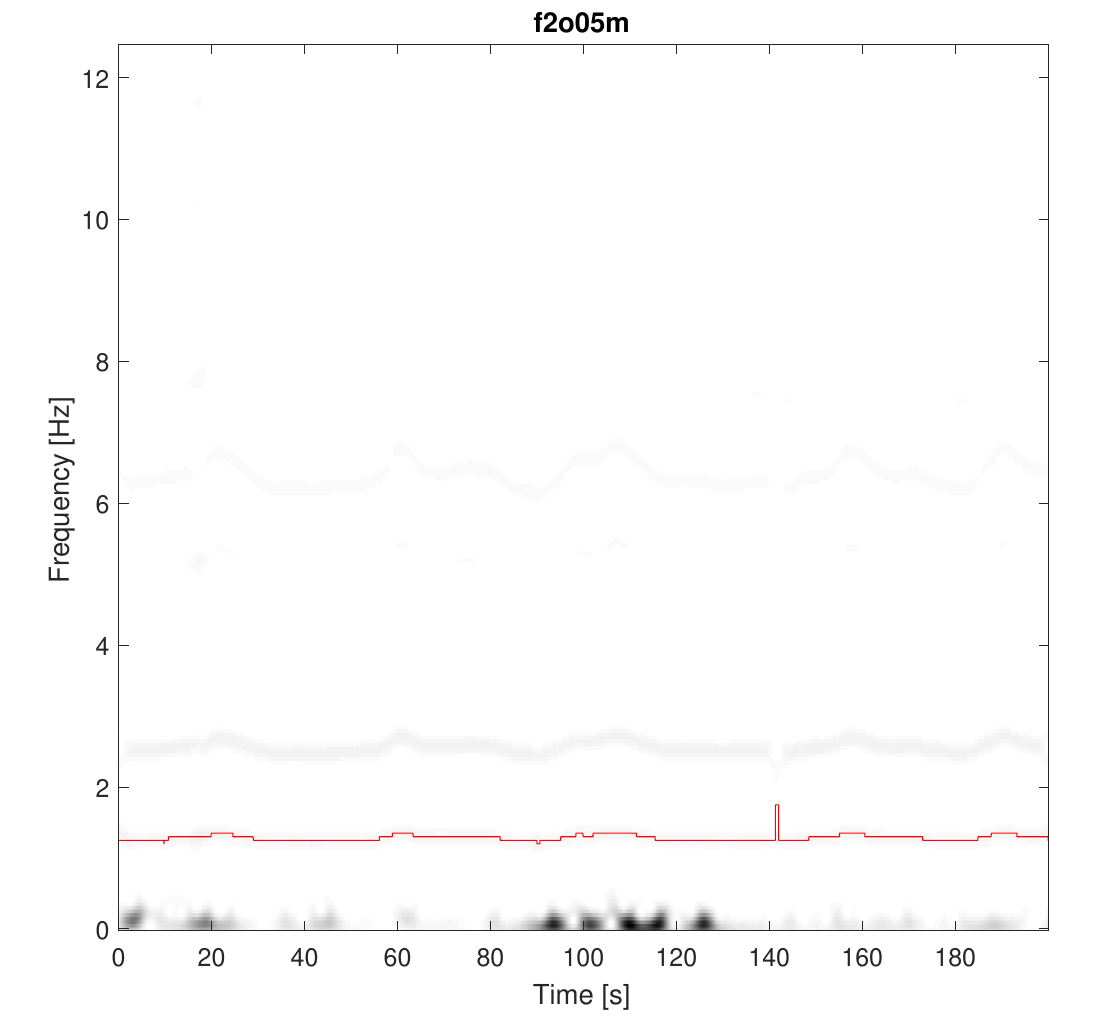}
	\includegraphics[width=0.45\textwidth]{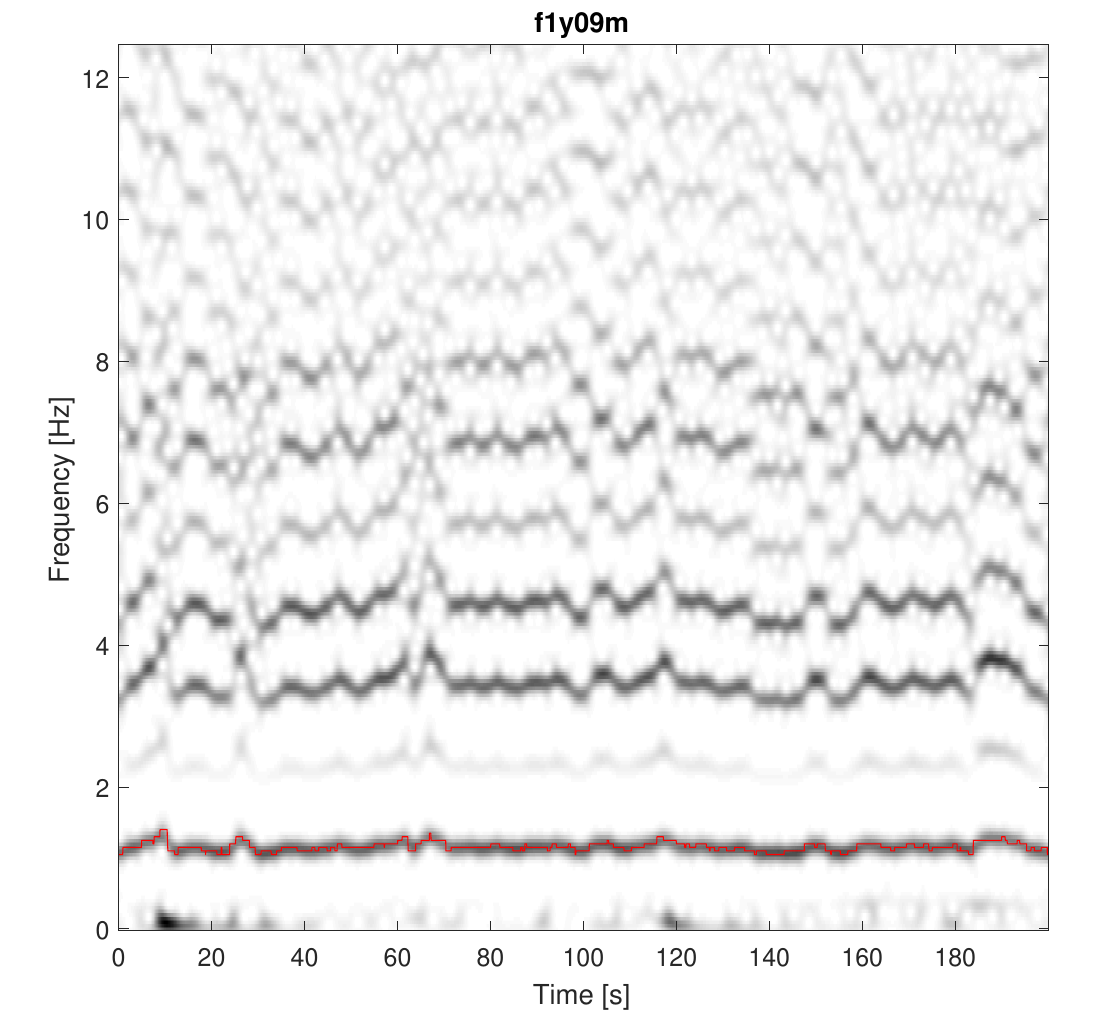}
	\caption{Spectrogram of the ECG signals for patients  f2o05m (left) and f1y09m (right) with the fundamental frequency ridges highlighted in red.}
	\label{fig:Spectro_Fantasia}
\end{figure}

This database was originally developed for the study of the dynamics responsible for the fluctuations in the heartbeat interval and its link to patient age~\cite{iyengar1996age}. It is comprised of 40 patients divided in two cohorts: 20 young patients (aged 21-34 years old) and 20 elderly patients (aged 68-85 years old). All patients underwent a rigorous screening procedure to discard potential health problems. Throughout the study the electrocardiogram (ECG) and respiratory signals were simultaneously recorded with the patient laying in supine position for 120 minutes while watching the movie \emph{Fantasia} (Disney, 1940) to maintain wakefulness. The recordings contain a considerable amount of noise and artifacts including baseline wandering and powerline interference.
We extracted $200$-seconds-long segments from these signals and applied the reconstruction procedure described in Sec. \ref{sec:ANH} using the generalized cross-validation criterion to choose the optimal model order $r$. In the ridge extraction step of the ECG signals, we computed the de-shape STFT~\cite{lin2018wave} to accurately extract the fundamental component of the signal (see the Supplemental Material for more details). We show the spectrograms of two signals from the database in Fig. \ref{fig:Spectro_Fantasia} where the fundamental ridges are highlighted in red. It becomes clear from the T-F representations that the presence of noise and baseline wandering makes it difficult to estimate the number of relevant harmonic components by individual ridge extraction. Additionally, these representations also show that the optimal number of harmonics that need to be considered for an accurate estimation of the WSF varies greatly from a patient to another. We estimate the WSFs and reconstruct the ECG and respiratory signals of each patient in both cohorts using Algo. 2 with automatic WSF order $r$ estimation. 

\begin{figure}[t]
	\centering
	\includegraphics[trim = 0 13 0 0,clip,width=0.45\textwidth]{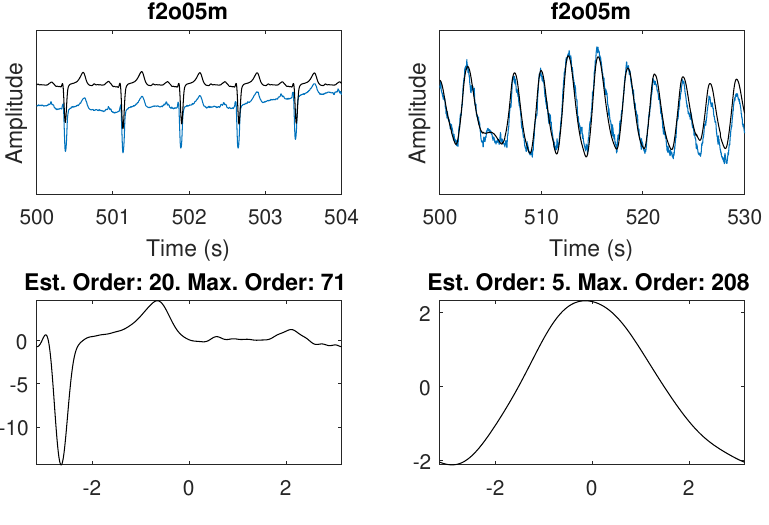}
	\includegraphics[width=0.45\textwidth]{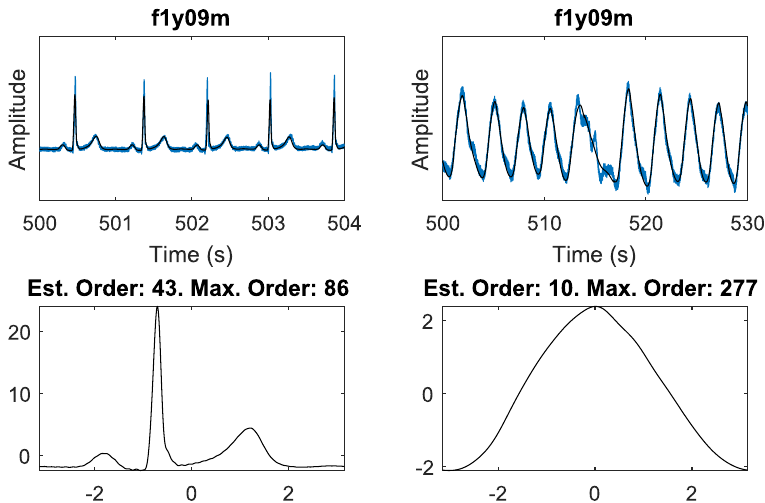}
	\caption{Left: WSF estimation and ANH reconstruction results for a sample patient from the elderly cohort of the Fantasia Database. The original signals are ploted in blue with the ANH model results superimposed in black. Right: Same signals from a sample patient of the young cohort.}
	\label{fig:FantasiaMain}
\end{figure}

In Fig. \ref{fig:FantasiaMain} we show the results for one patient from each age group. Alongside the estimated wave-shapes for the ECG and respiratory signals, we show the original signals in blue with the ANH model signal superimposed in black. We see that the ANH reconstruction recovers the general morphology of both the ECG and respiratory signals and filters the noise from both the powerline interference (f1y09m) and the baseline wandering (f2o05m). Additionally, it is worth noting that the optimal order for each signal is different and that, in general, more harmonics are needed to accurately represent the ECG as it is a more complex oscillatory phenomenon when compared to the respiratory signal. We indicate as `Est. Order' the optimal estimated order $r^*$, and `Max. Order', the maximum possible order $r_{max}$ according to the maximum fundamental frequency and the sampling rate. The same analysis was carried out for all patients in the database and the results can be found in the Supplemental Material.

\section{Conclusions}
In this paper, we proposed an algorithm for the automatic determination of the number of harmonics for the WSF model based on trigonometric regression model order estimation criteria. This allowed us to generate a fully adaptive signal reconstruction procedure. The performance of different order estimation criteria was evaluated for the case of non-stationary signals, with all criteria giving a good estimation of the model order under most conditions. Experimental results in simulated and real physiological signals show that the ANH model supported by the automatic order estimation procedure can accurately recover signals that are contaminated with considerable levels of noise, making this tool suitable for the task of decomposing and denoising biomedical signals.

\section*{Acknowledgements}
This work was supported jointly by CONICET and UNER through the Oriented Research Project PIO CONICET-UNER 14620140100014CO. The funding source is not involved in the completion of this work or the elaboration of the submitted manuscript.

\section*{Conflict of interest}

The authors declare that they have no conflict of interest.

\bibliography{references}

\begin{thebibliography}{10}
\expandafter\ifx\csname url\endcsname\relax
  \def\url#1{\texttt{#1}}\fi
\expandafter\ifx\csname urlprefix\endcsname\relax\def\urlprefix{URL }\fi
\expandafter\ifx\csname href\endcsname\relax
  \def\href#1#2{#2} \def\path#1{#1}\fi

\bibitem{flandrin2001}
P.~Flandrin, \href{https://doi.org/10.1117/12.421196}{{Time frequency and
  chirps}}, in: H.~H. Szu, D.~L. Donoho, A.~W. Lohmann, W.~J. Campbell, J.~R.
  Buss (Eds.), Wavelet Applications VIII, Vol. 4391, SPIE, 2001, pp. 161 --
  175.
\newblock \href {https://doi.org/10.1117/12.421196}
  {\path{doi:10.1117/12.421196}}.
\newline\urlprefix\url{https://doi.org/10.1117/12.421196}

\bibitem{picinbono1997instantaneous}
B.~Picinbono, On instantaneous amplitude and phase of signals, IEEE T. Signal
  Proces. 45~(3) (1997) 552--560.
\newblock \href {https://doi.org/10.1109/78.558469}
  {\path{doi:10.1109/78.558469}}.

\bibitem{wu2013instantaneous}
H.-T. Wu, Instantaneous frequency and wave shape functions (\uppercase{i}),
  Appl. Comput. Harmon. A. 35~(2) (2013) 181--199.
\newblock \href {https://doi.org/https://doi.org/10.1016/j.acha.2012.08.008}
  {\path{doi:https://doi.org/10.1016/j.acha.2012.08.008}}.

\bibitem{wu2014using}
H.-T. Wu, Y.-H. Chan, Y.-T. Lin, Y.-H. Yeh, Using synchrosqueezing transform to
  discover breathing dynamics from \uppercase{ECG} signals, Appl. Comput.
  Harmon. A. 36~(2) (2014) 354--359.
\newblock \href {https://doi.org/https://doi.org/10.1016/j.acha.2013.07.003}
  {\path{doi:https://doi.org/10.1016/j.acha.2013.07.003}}.

\bibitem{wu2016modeling}
H.-T. Wu, H.-K. Wu, C.-L. Wang, Y.-L. Yang, W.-H. Wu, T.-H. Tsai, H.-H. Chang,
  Modeling the pulse signal by wave-shape function and analyzing by
  synchrosqueezing transform, PloS one 11~(6) (2016).
\newblock \href {https://doi.org/https://doi.org/10.1371/journal.pone.0157135}
  {\path{doi:https://doi.org/10.1371/journal.pone.0157135}}.

\bibitem{su2017extract}
L.~Su, H.-T. Wu, Extract fetal \uppercase{ECG} from single-lead abdominal
  \uppercase{ECG} by de-shape short time \uppercase{F}ourier transform and
  nonlocal median, Front. App. M. 3 (2017) 2.
\newblock \href {https://doi.org/https://doi.org/10.3389/fams.2017.00002}
  {\path{doi:https://doi.org/10.3389/fams.2017.00002}}.

\bibitem{wu2014assess}
H.-t. Wu, R.~Talmon, Y.-L. Lo, Assess sleep stage by modern signal processing
  techniques, IEEE T. Bio-Med. Eng. 62~(4) (2014) 1159--1168.
\newblock \href {https://doi.org/10.1109/TBME.2014.2375292}
  {\path{doi:10.1109/TBME.2014.2375292}}.

\bibitem{cicone2017nonlinear}
A.~Cicone, H.-T. Wu, How nonlinear-type time-frequency analysis can help in
  sensing instantaneous heart rate and instantaneous respiratory rate from
  photoplethysmography in a reliable way, Front. Physiol. 8 (2017) 701.
\newblock \href {https://doi.org/https://doi.org/10.3389/fphys.2017.00701}
  {\path{doi:https://doi.org/10.3389/fphys.2017.00701}}.

\bibitem{sucic2011estimating}
V.~Sucic, N.~Saulig, B.~Boashash, Estimating the number of components of a
  multicomponent nonstationary signal using the short-term time-frequency
  r{\'e}nyi entropy, EURASIP Journal on Advances in Signal Processing 2011~(1)
  (2011) 1--11.

\bibitem{bruni2020signal}
V.~Bruni, M.~Tartaglione, D.~Vitulano, A signal complexity-based approach for
  am--fm signal modes counting, Mathematics 8~(12) (2020) 2170.

\bibitem{kavalieris1994determining}
L.~Kavalieris, E.~Hannan, Determining the number of terms in a trigonometric
  regression, J. Time Ser. Anal. 15~(6) (1994) 613--625.
\newblock \href
  {https://doi.org/https://doi.org/10.1111/j.1467-9892.1994.tb00216.x}
  {\path{doi:https://doi.org/10.1111/j.1467-9892.1994.tb00216.x}}.

\bibitem{meignen2012new}
S.~Meignen, T.~Oberlin, S.~McLaughlin, A new algorithm for multicomponent
  signals analysis based on synchrosqueezing: With an application to signal
  sampling and denoising, IEEE T. Signal Proces. 60~(11) (2012) 5787--5798.
\newblock \href {https://doi.org/10.1109/TSP.2012.2212891}
  {\path{doi:10.1109/TSP.2012.2212891}}.

\bibitem{colominas2020fully}
M.~A. Colominas, S.~Meignen, D.-H. Pham, Fully adaptive ridge detection based
  on stft phase information, IEEE Signal Proc. Let. 27 (2020) 620--624.
\newblock \href {https://doi.org/10.1109/LSP.2020.2987166}
  {\path{doi:10.1109/LSP.2020.2987166}}.

\bibitem{carmona1997characterization}
R.~A. Carmona, W.~L. Hwang, B.~Torr{\'e}sani, Characterization of signals by
  the ridges of their wavelet transforms, IEEE T. Signal Proces. 45~(10) (1997)
  2586--2590.
\newblock \href {https://doi.org/10.1109/78.640725}
  {\path{doi:10.1109/78.640725}}.

\bibitem{daubechies2011synchrosqueezed}
I.~Daubechies, J.~Lu, H.-T. Wu, Synchrosqueezed wavelet transforms: An
  empirical mode decomposition-like tool, Appl. Comput. Harmon. A. 30~(2)
  (2011) 243--261.
\newblock \href {https://doi.org/https://doi.org/10.1016/j.acha.2010.08.002}
  {\path{doi:https://doi.org/10.1016/j.acha.2010.08.002}}.

\bibitem{daubechies2016conceft}
I.~Daubechies, Y.~Wang, H.-T. Wu, Conce{FT}: Concentration of frequency and
  time via a multitapered synchrosqueezed transform, Philos. T. R. Soc. A
  374~(2065) (2016) 20150193.
\newblock \href {https://doi.org/https://doi.org/10.1098/rsta.2015.0193}
  {\path{doi:https://doi.org/10.1098/rsta.2015.0193}}.

\bibitem{kowalski2018convex}
M.~Kowalski, A.~Meynard, H.-T. Wu, Convex optimization approach to signals with
  fast varying instantaneous frequency, Appl. Comput. Harmon. A. 44~(1) (2018)
  89--122.
\newblock \href {https://doi.org/https://doi.org/10.1016/j.acha.2016.03.008}
  {\path{doi:https://doi.org/10.1016/j.acha.2016.03.008}}.

\bibitem{eubank1990curve}
R.~L. Eubank, P.~Speckman, Curve fitting by polynomial-trigonometric
  regression, Biometrika 77~(1) (1990) 1--9.
\newblock \href {https://doi.org/https://doi.org/10.1093/biomet/77.1.1}
  {\path{doi:https://doi.org/10.1093/biomet/77.1.1}}.

\bibitem{wax1985detection}
M.~Wax, T.~Kailath, Detection of signals by information theoretic criteria,
  IEEE T. Acoust. Speech 33~(2) (1985) 387--392.
\newblock \href {https://doi.org/10.1109/TASSP.1985.1164557}
  {\path{doi:10.1109/TASSP.1985.1164557}}.

\bibitem{akaike1998information}
H.~Akaike, Information theory and an extension of the maximum likelihood
  principle, in: Selected papers of Hirotugu Akaike, Springer, 1998, pp.
  199--213.

\bibitem{rissanen1978modeling}
J.~Rissanen, Modeling by shortest data description, Automatica 14~(5) (1978)
  465--471.
\newblock \href {https://doi.org/https://doi.org/10.1016/0005-1098(78)90005-5}
  {\path{doi:https://doi.org/10.1016/0005-1098(78)90005-5}}.

\bibitem{wang1993aic}
X.~Wang, An \uppercase{AIC} type estimator for the number of cosinusoids, J.
  Time Ser. Anal. 14~(4) (1993) 433--440.
\newblock \href
  {https://doi.org/https://doi.org/10.1111/j.1467-9892.1993.tb00155.x}
  {\path{doi:https://doi.org/10.1111/j.1467-9892.1993.tb00155.x}}.

\bibitem{quinn1989estimating}
B.~G. Quinn, Estimating the number of terms in a sinusoidal regression, J. Time
  Ser. Anal. 10~(1) (1989) 71--75.
\newblock \href
  {https://doi.org/https://doi.org/10.1111/j.1467-9892.1989.tb00016.x}
  {\path{doi:https://doi.org/10.1111/j.1467-9892.1989.tb00016.x}}.

\bibitem{hannan1984multivariate}
E.~J. Hannan, L.~Kavalieris, Multivariate linear time series models, Adv. Appl.
  Probab. (1984) 492--561\href
  {https://doi.org/https://doi.org/10.2307/1427286}
  {\path{doi:https://doi.org/10.2307/1427286}}.

\bibitem{donoho1994ideal}
D.~L. Donoho, J.~M. Johnstone, Ideal spatial adaptation by wavelet shrinkage,
  Biometrika 81~(3) (1994) 425--455.
\newblock \href {https://doi.org/https://doi.org/10.1093/biomet/81.3.425}
  {\path{doi:https://doi.org/10.1093/biomet/81.3.425}}.

\bibitem{pham2018novel}
D.-H. Pham, S.~Meignen, A novel thresholding technique for the denoising of
  multicomponent signals, in: 2018 Int. Conf. Acoust. Spee., IEEE, 2018, pp.
  4004--4008.
\newblock \href {https://doi.org/10.1109/ICASSP.2018.8462216}
  {\path{doi:10.1109/ICASSP.2018.8462216}}.

\bibitem{lin2018wave}
C.-Y. Lin, L.~Su, H.-T. Wu, Wave-shape function analysis, J. Fourier Anal.
  Appl. 24~(2) (2018) 451--505.
\newblock \href {https://doi.org/10.1007/s00041-017-9523-0}
  {\path{doi:10.1007/s00041-017-9523-0}}.

\bibitem{pwdb2019}
P.~H. Charlton, J.~M. Harana, S.~Vennin, Y.~Li, P.~Chowienczyk, J.~Alastruey,
  \href{https://zenodo.org/record/3275625\#.YSk9xt-vHIU}{Pulse wave database
  (\uppercase{PWDB}): A database of arterial pulse waves representative of
  healthy adults, version 0.2.0, zenodo} (2019).
\newblock \href {https://doi.org/10.5281/zenodo.3275625}
  {\path{doi:10.5281/zenodo.3275625}}.
\newline\urlprefix\url{https://zenodo.org/record/3275625\#.YSk9xt-vHIU}

\bibitem{charlton2019modeling}
P.~H. Charlton, J.~Mariscal~Harana, S.~Vennin, Y.~Li, P.~Chowienczyk,
  J.~Alastruey, Modeling arterial pulse waves in healthy aging: a database for
  in silico evaluation of hemodynamics and pulse wave indexes, Am. J.
  Physiol.-Heart C. 317~(5) (2019) H1062--H1085.
\newblock \href {https://doi.org/https://doi.org/10.1152/ajpheart.00218.2019}
  {\path{doi:https://doi.org/10.1152/ajpheart.00218.2019}}.

\bibitem{kamen1996poincare}
P.~W. Kamen, H.~Krum, A.~M. Tonkin, Poincare plot of heart rate variability
  allows quantitative display of parasympathetic nervous activity in humans,
  Clin. Sci. 91~(2) (1996) 201--208.
\newblock \href {https://doi.org/10.1042/cs0910201}
  {\path{doi:10.1042/cs0910201}}.

\bibitem{Baraniuk2001}
R.~G. Baraniuk, P.~Flandrin, A.~J. Janssen, O.~J. Michel, Measuring
  time-frequency information content using the {R}{\'e}nyi entropies, IEEE T.
  Inform. Theory 47~(4) (2001) 1391--1409.
\newblock \href {https://doi.org/10.1109/18.923723}
  {\path{doi:10.1109/18.923723}}.

\bibitem{meignen2020use}
S.~Meignen, M.~A. Colominas, D.-H. Pham, On the use of {R}{\'e}nyi entropy for
  optimal window size computation in the short-time {F}ourier transform, in:
  2020 Int. Conf. Acoust. Spee., IEEE, 2020, pp. 5830--5834.
\newblock \href {https://doi.org/10.1109/ICASSP40776.2020.9053392}
  {\path{doi:10.1109/ICASSP40776.2020.9053392}}.

\bibitem{fantasia2003}
C.-K. Peng, L.~Lipsitz,
  \href{https://physionet.org/content/fantasia/1.0.0/}{Fantasia database, v1,
  physionet} (2003).
\newblock \href {https://doi.org/https://doi.org/10.13026/C2RG61}
  {\path{doi:https://doi.org/10.13026/C2RG61}}.
\newline\urlprefix\url{https://physionet.org/content/fantasia/1.0.0/}

\bibitem{goldberger2000physiobank}
A.~L. Goldberger, L.~A. Amaral, L.~Glass, J.~M. Hausdorff, P.~C. Ivanov, R.~G.
  Mark, J.~E. Mietus, G.~B. Moody, C.-K. Peng, H.~E. Stanley, Physiobank,
  physiotoolkit, and physionet: components of a new research resource for
  complex physiologic signals, Circulation 101~(23) (2000) e215--e220.
\newblock \href {https://doi.org/10.1161/01.cir.101.23.e215}
  {\path{doi:10.1161/01.cir.101.23.e215}}.

\bibitem{iyengar1996age}
N.~Iyengar, C.~Peng, R.~Morin, A.~L. Goldberger, L.~A. Lipsitz, Age-related
  alterations in the fractal scaling of cardiac interbeat interval dynamics,
  Am. J. Physiol.-Reg. I. 271~(4) (1996) R1078--R1084.
\newblock \href
  {https://doi.org/https://doi.org/10.1152/ajpregu.1996.271.4.R1078}
  {\path{doi:https://doi.org/10.1152/ajpregu.1996.271.4.R1078}}.

\end{thebibliography}
\bibliographystyle{elsarticle-num}

\begin{center}
\begin{Large}
	 Supplemental Material for Wave-shape Function Model Order Estimation by Trigonometric Regression
	\end{Large}
\end{center}
%\vspace
%\newpage
	
%\title{:}
	
%\maketitle

\renewcommand{\thefigure}{S.\arabic{figure}}
\renewcommand{\theequation}{S.\arabic{equation}}

We present in this supplemental document the complete results for the adaptive wave-shape estimation experiment performed on electrocardiography (ECG) and respiratory signals from the Fantasia Database. This document is structured as follows. In Sec. S1, we describe the Fantasia Database. In Sec. S2, we give the details of the signal analysis and wave-shape estimation procedure. Finally, the results of the experiment are discussed in Sec. S3. 

\setcounter{section}{0}
\setcounter{figure}{0}
\section{Fantasia Database}
The Fantasia Database was originally developed for the study of fluctuations in the cardiac interbeat interval of healthy subjects. The objective of this study was to better comprehend the underlying dynamics of the heartbeat, which is known to have a fractal like structure~\cite{iyengar1996age}. This database is made available through the Physionet website~\cite{goldberger2000physiobank}.

The Database is comprised of the records from 40 patients divided in two cohorts: 20 young patients (aged 21-34 years old) and 20 elderly patients (aged 68-85). All patients underwent a rigorous screening procedure in order to discard potential health problems. This screening included physical examination, biochemical blood analysis, ECG, and exercise tolerance test. Throughout the study, the electrocardiogram (ECG) and respiratory signals are simultaneously recorded with the patient laying in supine position for 120 minutes while watching the movie \emph{Fantasia} (Disney, 1940) to maintain wakefulness. In half of each group, the uncalibrated non-invasive blood pressure signal is also recorded.

All recorded signals were digitized at a sampling rate of 250 Hz. As shown in Figs. \ref{fig:Fantasia1}-\ref{fig:Fantasia4} all signals contain considerable noise and artifacts, including powerline interference (50 Hz) and baseline wandering. 
\section{Adaptive Wave-Shape Estimation in Physiological Signals}

We perform the following experiment to show that a fully adaptive wave-shape function estimation, including automatic model order selection, is needed in order to accurately reconstruct the ECG and respiratory signals while taking into account interpatient waveform variability. 
For each subject, a 200 seconds-long segment is extracted from both the ECG and respiratory signal. We exclude the blood pressure signals from this study since it is only available for half of the subjects in the database. The estimation of the fundamental instantaneous frequency for each segment is performed using the ridge extraction procedure described in Algo. 1. In the case of the ECG recordings, the fundamental component is rarely the dominant ridge in the T-F domain due to the complex morphology of the wave. Therefore, the ridge detection based on maximizing the energy of the spectrogram cannot recover the fundamental instantaneous frequency of the signal. In order to overcome this problem, we apply the de-shape STFT technique proposed by Lin et al~\cite{lin2018wave}. Briefly, we compute the standard STFT $F(t,f)$ and then derive the de-shape STFT $W(t,f)$ by the following operation: 
\begin{equation}
	W(t,f) = F(t,f)U(t,f),
\end{equation}

\noindent where $U(t,f)$ is obtained by inverting the quefrency axis of the so called short-time cepstrum transform (STCT):
\begin{align}
	U(t,f)&=C(t,1/q)\\
	C(t,q)&=\int |F(t,f)|^\gamma e^{-2\pi i q f} df.
\end{align}

For the following experiments, we set the exponent $\gamma$ of the root function $g_\gamma(x) = |x|^\gamma$ is set as $\gamma = 0.3$. Because the spectral information of the STFT and the inverted STCT overlap only in the region of the fundamental component of the signal, the dominant ridge of the de-shape STFT is located in this region. Therefore, we can obtain the fundamental ridge by replacing $F(t,f)$ for $W(t,f)$ in the ridge extraction procedure shown in Step 4 of Algo. 1 (main document). We then estimate the instantaneous amplitude and phase from the fundamental component of the signal from the STFT (using the frequency ridge extracted from the de-shape STFT) and use these estimates to construct the pseudo-Fourier dictionary $\mathbf{C_r}$ needed to solve the least square problem in Eq. \ref{eq:regression}. As mentioned in Sec. \ref{sec:ANH}, we need to chose a value for the parameter $r$, the number of terms of the Fourier expansion of the WSF. We apply the order estimation method based on trigonometric model selection criteria to estimate this parameter. For each signal segment, we find the WSF optimal order as

\begin{equation}
	r^*=\underset{r\in[1,r_{max}]}{\text{argmax}}\Phi_G(r),
\end{equation}

\noindent where $\Phi_G(r)$ is the generalized cross-validation criterion as defined in Sec. \ref{sec:Criteria_RSS}. Finally, we reconstruct the signal using the optimal number of terms $r^*$.

\section{Experimental Results}
\begin{figure*}
	\centering
	\includegraphics[width=\textwidth]{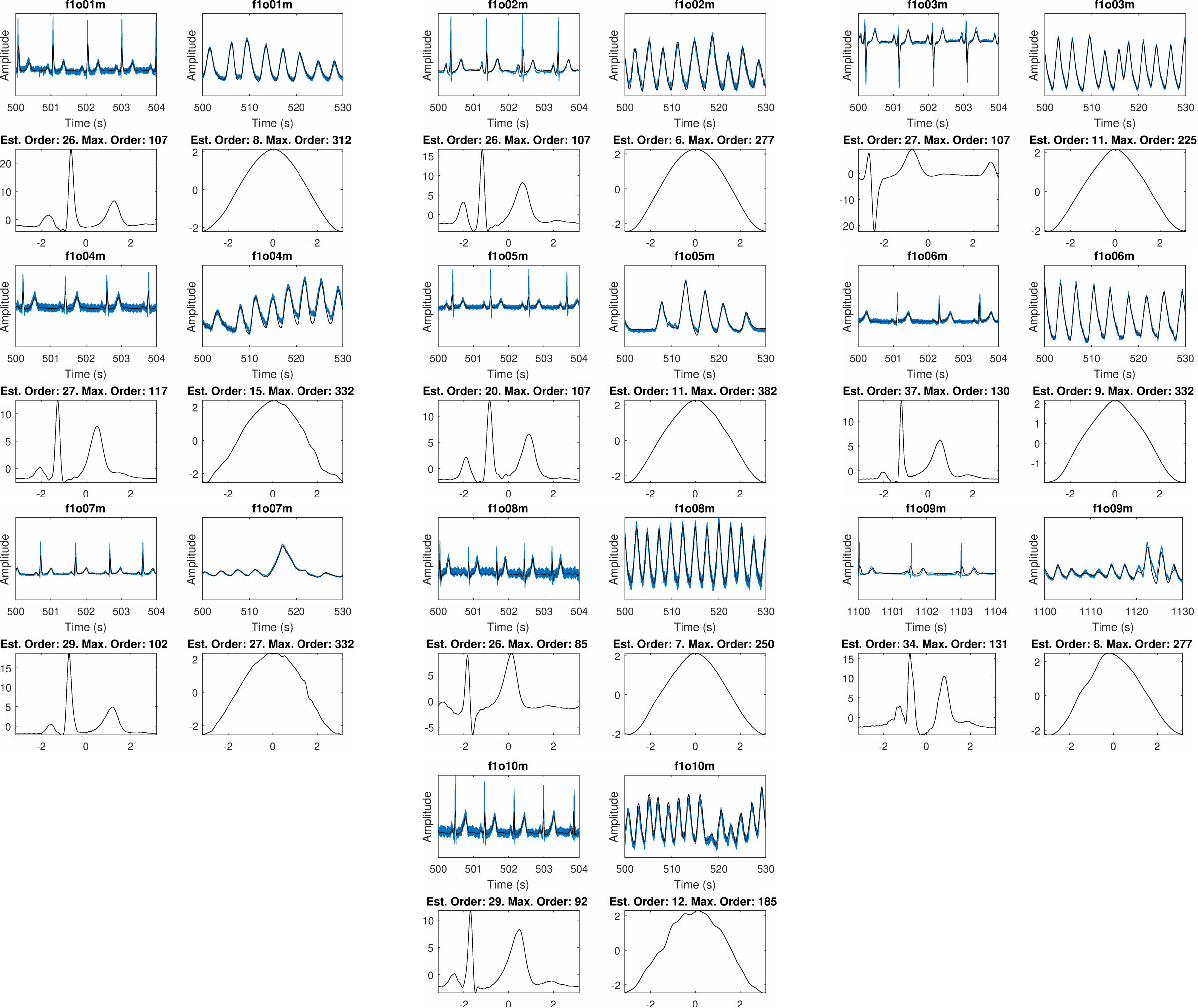}
	\caption{Results for the first 10 subjects in the elderly cohort. Original ECG and respiratory signals are plotted in blue with the reconstructed signal superimposed. WSFs for each signal are also shown.}
	\label{fig:Fantasia1}
\end{figure*}

\begin{figure*}
	\centering
	\includegraphics[width=\textwidth]{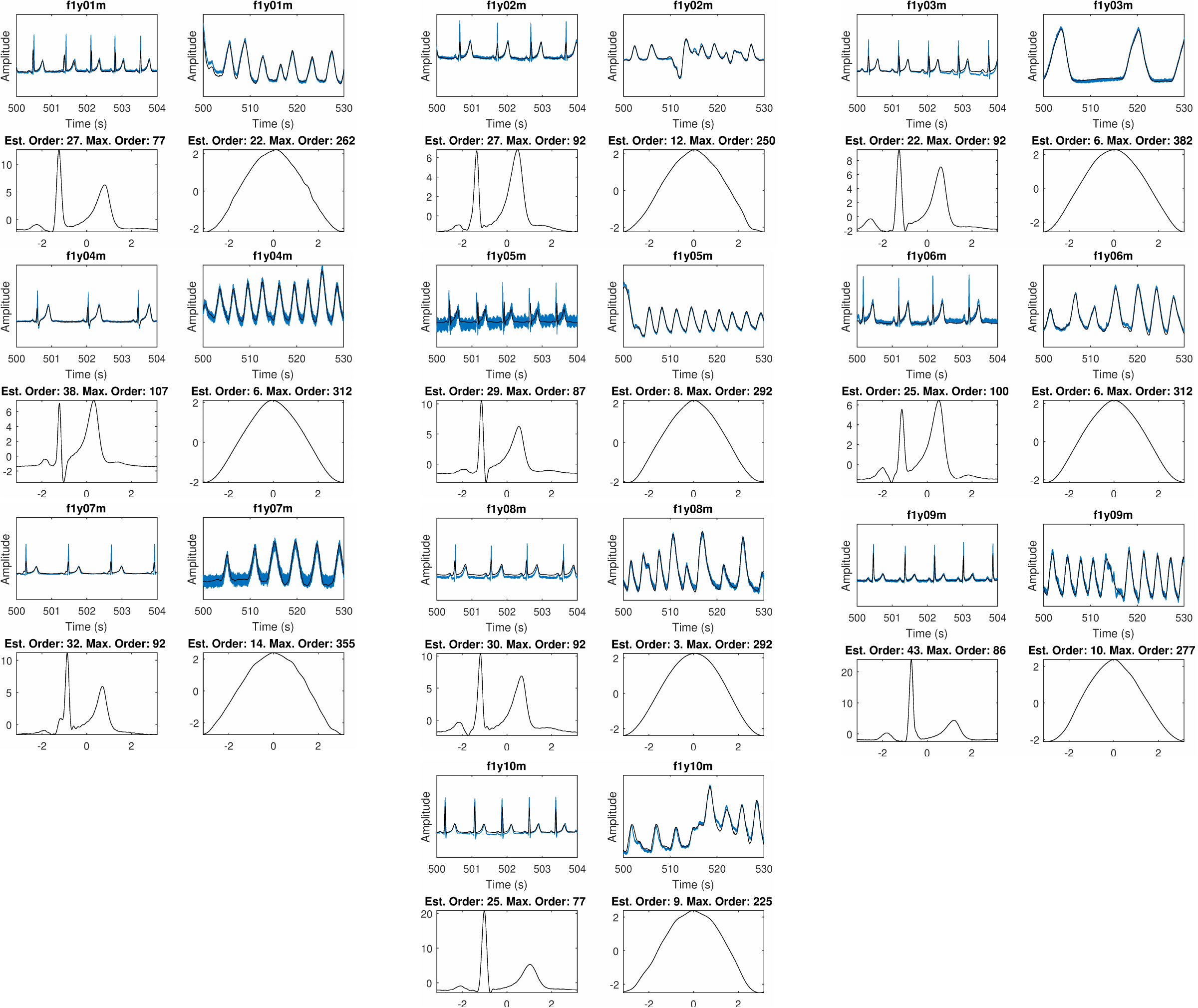}
	\caption{Results for the first 10 subjects in the young cohort. Original ECG and respiratory signals are plotted in blue with the reconstructed signal superimposed. WSFs for each signal are also shown.}
	\label{fig:Fantasia2}
\end{figure*}

\begin{figure*}
	\centering
	\includegraphics[width=\textwidth]{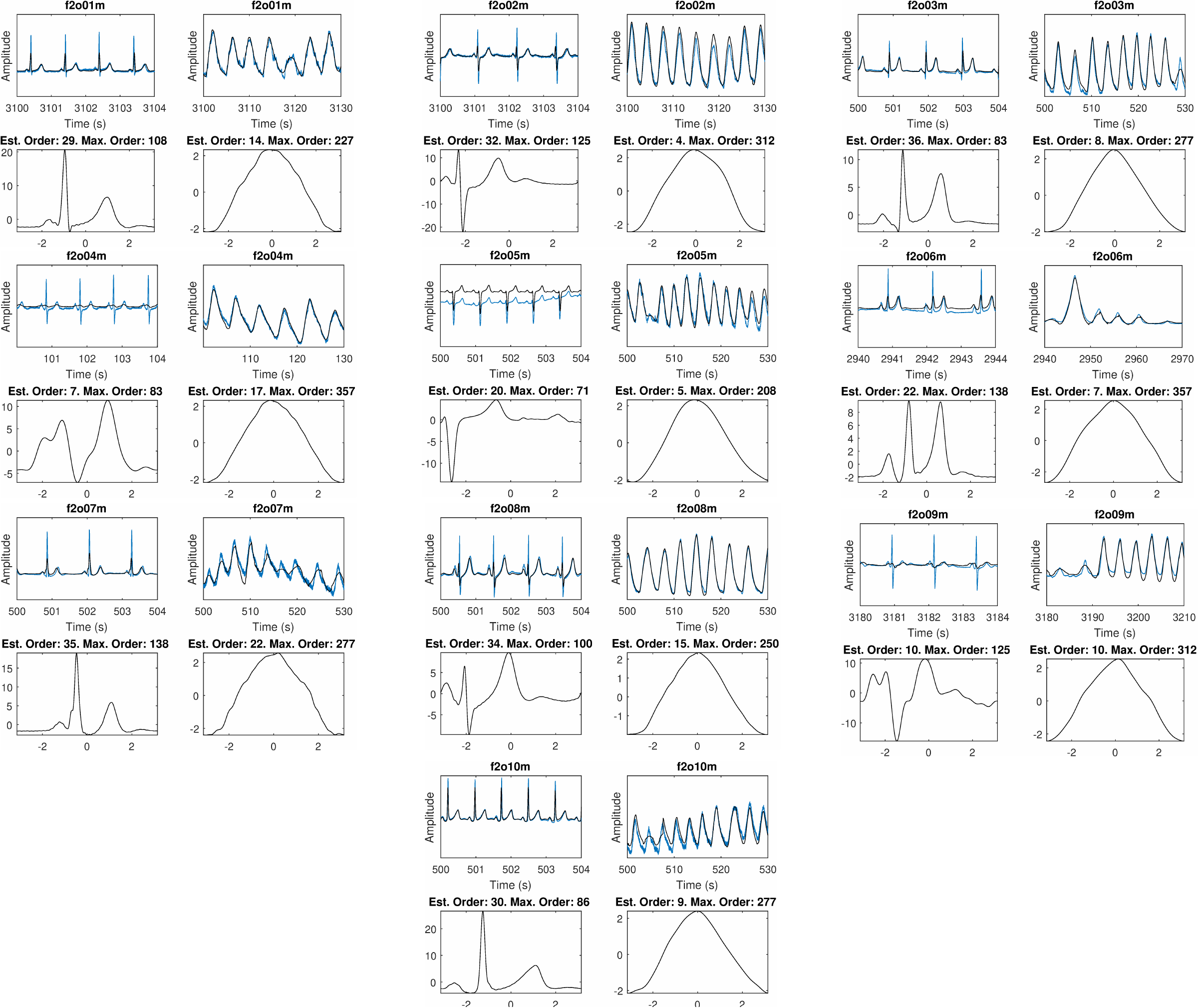}
	\caption{Results for the last 10 subjects in the elderly cohort. Original ECG and respiratory signals are plotted in blue with the reconstructed signal superimposed. WSFs for each signal are also shown.}
	\label{fig:Fantasia3}
\end{figure*}

\begin{figure*}
	\centering
	\includegraphics[width=\textwidth]{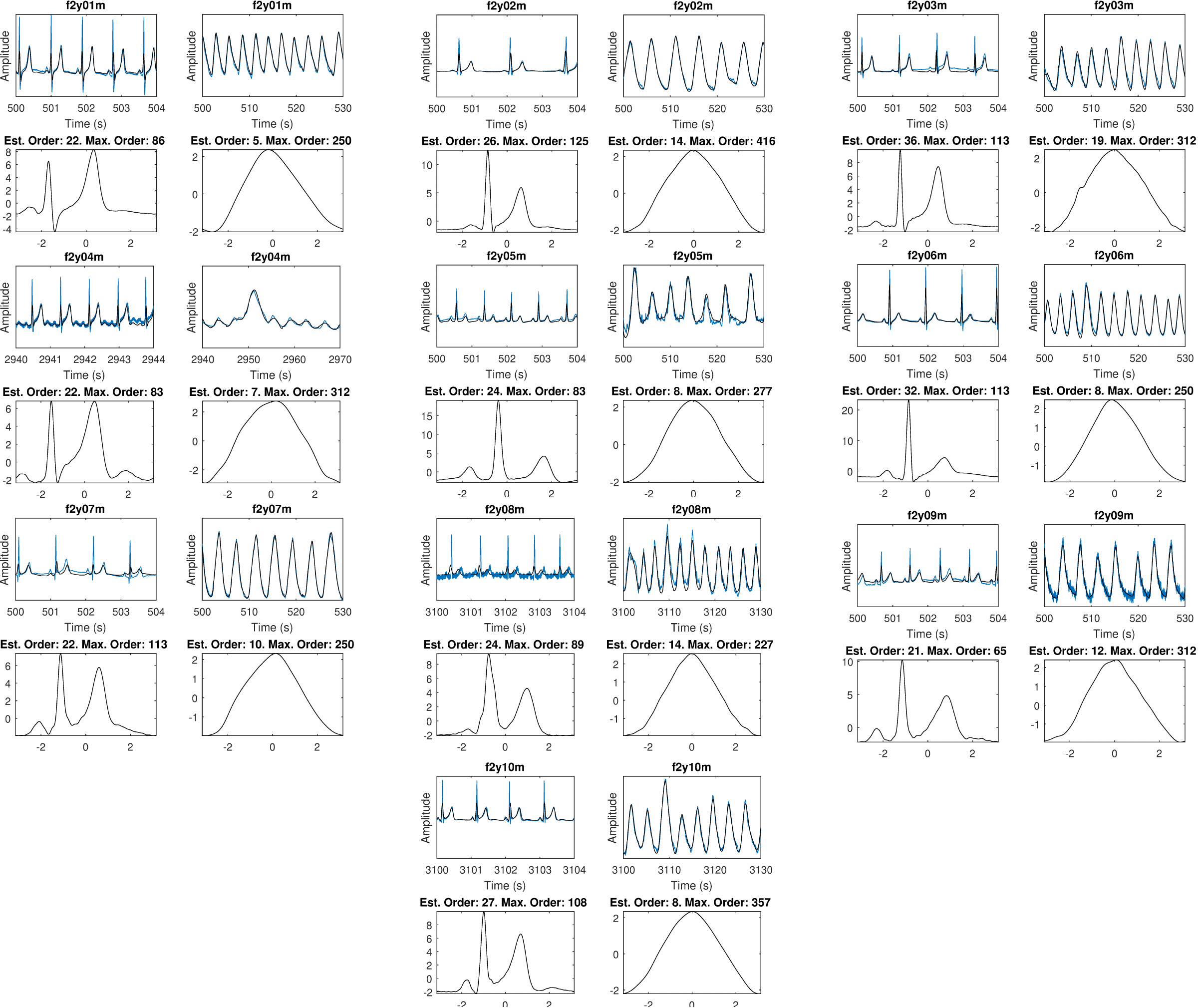}
	\caption{Results for the last 10 subjects in the young cohort. Original ECG and respiratory signals are plotted in blue with the reconstructed signal superimposed. WSFs for each signal are also shown.}
	\label{fig:Fantasia4}
\end{figure*}

We show the results of this experiment in Figs. \ref{fig:Fantasia1}-\ref{fig:Fantasia4}. For both the ECG and respiratory signal of each patient, the original (noisy) signal is plotted in blue with the reconstructed signal superimposed in black. We also include the estimated WSF for each signal in the range $(-\pi,\pi)$. We see that the reconstruction procedure based on the ANH model accurately captures the oscillatory pattern for both ECG and respiration while also eliminating the powerline interference and baseline wandering present in the signals. One of the negative aspects of the reconstruction is that the R peak in the ECG is underestimated in most cases, although the degree of R peak attenuation is not consistent for all signals. In some, the amplitude of the T wave surpasses that of the R peak. As we can see from the wave-shapes in Figs. \ref{fig:Fantasia1}-\ref{fig:Fantasia4}, the optimal number of trigonometric terms obtained based on the selection criterion are different for most patients and noticeably low when compared to the maximum order admissible $r_{max}$. Respiratory signals have a simpler waveform, so lower model order are favoured. In contrast, the morphology of the ECG is more complex and varies greatly from one patient to another, which requires a higher order model. Particularly, for respiratory signals, the ANH model reconstruction is able to preserve the rapid changes in both instantaneous frequency and amplitude due to the nature of the physiological respiratory phenomenon and the signal acquisition conditions. Nevertheless, the reconstruction procedure fails in some cases. For the f2o04m patient the ANH model is unable to recover the QRS complex of the ECG and the amplitude of both the P and T waves is considerably reduced. For the f2o09m patient the reconstruction procedure also fails to capture the QRS complex and the optimal order of the WSF is particularly low ($10$ harmonic components). 

We surmise that the reason for this poor performance is that the low-frequency noise overlaps with the fundamental component region in the T-F plane, degrading the estimators for the fundamental amplitude and phase of the signal.  Notwithstanding the above, the results we obtained indicate that the automatic order estimation using trigonometric model selection criteria is useful for the modelling of real physiological signals that have time-varying properties (phase and amplitude) in the presence of noise. Another possible cause of errors in the approximation is that the low sampling frequency cannot capture the information of the QRS complex with enough resolution for it to be modelled adequately by a finite sum of trigonometric terms.

\newcommand\w{0.3}

\end{document}